\documentclass[]{jfm}

\usepackage[dvipsnames]{xcolor}
\usepackage{comment}
\usepackage{graphicx}
\usepackage{newtxtext}
\usepackage{newtxmath}
\usepackage{natbib}
\usepackage{hyperref}
\usepackage{gensymb}
\hypersetup{
    colorlinks = true,
    urlcolor   = blue,
    citecolor  = black,
}

\newcommand{\RomanNumeralCaps}[1]

\usepackage{graphicx}
\usepackage{epstopdf, epsfig}
\usepackage{amsmath}
\usepackage{array}
\usepackage{multirow}
\usepackage[utf8]{inputenc}

\usepackage{subfigure}

\usepackage{pgfplots}
\usepackage{tikz}
\usetikzlibrary{backgrounds,pgfplots.groupplots,external}
\newcommand{\pd}[2]{\frac{\partial #1}{\partial #2}}

\title{Releasing trapped Taylor bubbles via centrifugation and inclination}

\author{Alice Marcotte \aff{1}, Pier Giuseppe Ledda \aff{2}, Valentin Buriasco\aff{3}, Paul Dené\aff{3}, François Gallaire \aff{3}, Ludovic Keiser \aff{4}}

\affiliation{\aff{1} Institut Jean le Rond d'Alembert, Sorbonne Université, Paris, France, \aff{2} Dipartimento di Ingegneria Civile, Ambientale e Architettura, Università degli Studi di Cagliari, Cagliari, Italy, \aff{3} Laboratory of Fluid Mechanics and Instabilities, École Polytechnique Fédérale de Lausanne, Lausanne, Switzerland, \aff{4} Institut de Physique de Nice, Université Côte d'Azur, Nice, France}

\begin{document}
\maketitle

\begin{abstract}
The entrainment of air in liquid-filled channels  occurs across diverse applications, such as CO$_2$ sequestration, circulatory biological systems, and microfluidics.
In confined systems, the entrapment of a gas volume with an equivalent spherical diameter greater than the dimension of the channel, results in the formation of an extended bubble that can obstruct the fluid circuit and eventually compromise performance. Notably, in sealed vertical tubes, buoyant long bubbles - called Taylor bubbles - cannot rise if the inner tube radius is lower than a critical value close to the capillary length. This critical threshold for steady ascent has been shown to be determined by geometrical constraints related to the required matching of the upper cap shape with the lubricating film developing in the elongated part of the bubble.

As a matter of fact, in application fields involving narrow liquid channels, long bubbles may be challenging
to eliminate. In this context, developing strategies to overcome the motion threshold and release stuck bubbles
is desirable. Such strategies require to modify the matching conditions by means of an external force field, in a
way that favours bubble ascent. However, it remains unclear how changes in acceleration conditions affect the onset of motion of buoyancy-driven long bubbles.

This study investigates the onset of motion and the resulting velocity modulation beyond threshold of elongated bubbles in sealed tubes with an inner radius near the critical value, where bubble motion is inhibited in a vertical setting. Two strategies are explored to tune bubble motion, which exploit variations of the axial and transversal gravitational accelerations: tube rotation around its symmetry axis and tube inclination with respect to gravity. 
By revising and extending the matching conditions and the resulting geometrical constraints of the simple vertical setting,
the study predicts the new thresholds based on rotational speed and tilt angle, respectively, providing predictions for the rising velocity of bubbles under the resulting modified apparent gravity. Experimental measurements of the motion threshold and rising velocity are compared against our theoretical developments, with a good agreement, thus offering practical approaches to control and tune bubble motion in confined environments.

\end{abstract}

\section{Introduction}

Air entrapment into liquid-filled channels is encountered in a broad range of applications, from simple hydraulic systems for intravenous filling \citep{Groell1997} to CO$_2$ sequestration in depleted geological oil reservoirs \citep{Oldenburg2006, Corapcioglu2004,Wang2012}, embolism in circulatory biological systems \citep{Brodribb2016,Li2021}, and multiphase microfluidics flows \citep{Baroud2010}. 
In miniaturized fluid systems, air bubbles can be exploited for transport of particles or for mixing processes \citep{baroud2007optical,Baroud2010,citStoneReview}. Conversely, long gas bubbles may represent a challenging issue, since they can occlude the entire cross-section of the channel and reduce the performance of the fluid circuit \citep{Jensen2004, vanSteijn2008, Brodribb2016}. 

In application fields involving narrow liquid channels, long bubbles may be challenging to eliminate, while 
disrupting fluid flow, causing pressure fluctuations and affecting mixing processes.  
In perfusion systems for cell cultures, these bubbles can have several detrimental effects on cell health and experimental outcomes, such as localized nutrient deprivation, altered pH levels, and accumulation of waste products, all of which can negatively impact cell viability and function \citep{Sung2009}. As an additional example, in fuel cells, the oxidation of methanol leads to the formation of CO$_2$ bubbles, that reduce the cell's efficiency \citep{Litterst2006}. Thus, a considerable effort has been dedicated to the removal of bubbles in these circuits (see among others \cite{Sung2009}, \cite{Cheng2014}, \cite{Guo2022}).

Conversely, transport of long bubbles in microfluidic channels can be cleverly exploited, for instance for particle sieving. Since the bubble speed is intrinsically linked to the thickness of its surrounding lubricating film, tuning the velocity of the bubble may be used to separate particles based on their size \citep{Yu2018}: monitoring the speed of the bubble may prevent particles larger than the film thickness to reach the fluid region past
the bubble. The bubble thus acts as an active filter that has the high advantage of preventing clogging. Thus, enabling and controlling the motion of elongated bubbles in capillaries can enhance the efficiency of these microfluidic systems. 

\begin{figure}
\centering
\includegraphics[width=\textwidth]{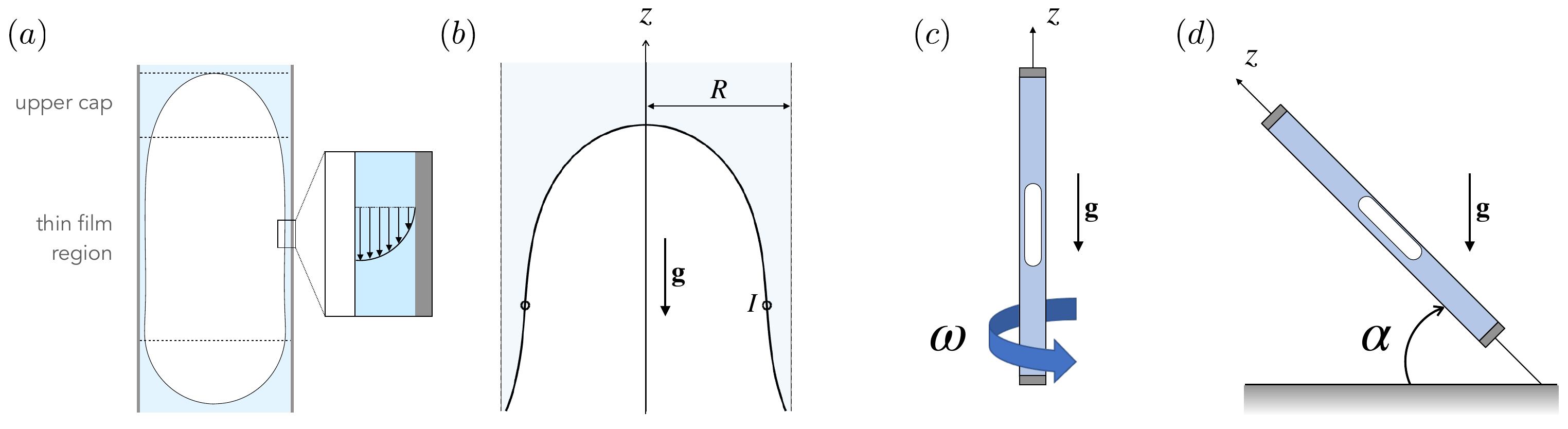}
\caption{(a) Schematics of a long bubble immersed in a viscous liquid inside a sealed capillary. The top part of the bubble can be divided into an upper cap and into an elongated part surrounded by a thin film. For a buoyant bubble to rise, mass conservation requires  the fluid displaced by the tip of the bubble to drain through the thin film. (b) Sketch of the upper cap profile of a long air bubble within a sealed tube of radius $R$, in the vertical setting studied by \cite{Bretherton1961}. The profile exhibits an inflection point denoted by $I$. For $R>R_c$, the matching with the thin film region at the inflection point is possible. (c) and (d) Sketch of the configurations investigated in this study. In the first case (c), the tube is hold vertically and rotates around its symmetry axis at angular frequency $\omega$. In the second case (d), the tube is tilted with respect to gravity and makes an angle $\alpha$ with the horizontal plane. }
\label{fig:intro}

\end{figure}

Many hydraulic and microfluidic systems rely on vertical settings (see, for instance, \citealp{kaigala2011vertical}), thereby calling for a better understanding of how bubble transport is influenced by gravity forces. In a vertical configuration, a gas volume in a liquid-filled channel is expected to rise owing to buoyancy.
The more specific case of buoyant ascent in a vertical tube of a gas volume with an equivalent spherical diameter larger than the tube inner radius, has been investigated in the seminal work by \cite{Davies1950}, that provided a prediction for the rising velocity of long bubbles in tubes, subsequently termed as Taylor bubbles. 
However, it was observed over a century ago \citep{Gibson1913} that long bubbles within a sealed vertical tube with a sufficiently narrow diameter exhibit an interesting behavior: they cease to rise and appear to be stuck. \cite{Bretherton1961} showed that if the tube inner radius was smaller than a critical value $R_c$ close to the capillary length $\ell_c$ of the liquid (more precisely,  $R_c \approx 0.918 \ell_c$), no valid bubble shape was compatible with a steady rising motion. This threshold stems from the asymptotic matching between the upper cap profile, which results from the equilibrium between surface tension and gravity, and the thin film surrounding the elongated part of the bubble, where viscous, surface tension and gravity forces are balanced. These two regions are depicted in Figure \ref{fig:intro}(a). The condition for the onset of motion can be summarized as a geometrical constraint which imposes, for the existence of a steadily ascending bubble, that the upper cap profile exhibits an inflection point with negative slope (for an upward oriented vertical axis), see Figure \ref{fig:intro}(b).  At the critical condition $R=R_c$, both the slope and curvature vanish at the solid wall. In addition, for $R$ slightly larger than $R_c$, \cite{Bretherton1961} predicted the bubble rising velocity, by exploiting mass conservation through the thin film and the variation of the slope at the inflection point with the tube's radius.

Below the threshold  $R<R_c$, \cite{Lamstaes2017} studied the unsteady bubble motion and predicted the occurrence of a self-similar pinch-off singularity of the thin lubricating film around the bubble, thus hindering any further flow and eventually stopping the progression of the bubble, found to travel a finite distance over infinite time. That prediction is supported by recent interference microscopy experiments, that have demonstrated that the bubble is apparently stuck by an infinitely slow flow taking place in the surrounding thin liquid film whose nanometric thickness results from an equilibrium between capillary stress and disjoining pressure \citep{Dhaouadi2019}. 

For bubble ascent in sealed tubes, it is necessary for the fluid displaced by the tip of the bubble to drain through the thin film. Thus, enabling the motion of the bubble in sealed tubes with inner radii smaller than the critical value $R_c$, requires to develop some strategies that would act on the thickness of the surrounding lubricating film. \cite{Zhou2021} showed numerically that "encaging" the bubble by means of thin vertical rods regularly arranged on a circle coaxial with the tube could effectively expand the gap between the air-liquid interface and the inner solid wall, thus facilitating the downward flow of the liquid and increasing the rising velocity of the bubble. \cite{Bi2001} and \cite{Bico2002} demonstrated that using angular tubes could effectively promote the rising of the bubble even under strong confinement, owing to the presence of corners that allow for a more efficient drainage of the liquid around the bubble \citep{Funada2005}. In the same spirit, another strategy consists in using textured inner walls: because of the imbibition of the roughness, the effective thickness of the lubricating film is actually larger than on a smooth surface \citep{Bico2001}. 

However, in some applications where the geometry of the tube cannot be modified adequately, the film thickness could be varied by adjusting the pressure distribution in the surrounding liquid by mean of an external force field, which could be easily tuned so as to precisely control the ascent velocity of the bubble. In this context, it has been shown that imposing a liquid flow in the tube effectively thickens the lubricating film around the bubble \citep{Yu2021}. In particular, \cite{Magnini2019} demonstrated that when the external flow is oriented in the same (upward) direction as buoyancy, it can enable the rise of bubbles in tubes with radius $R<R_c$. \cite{Kubie2000} documented a significant increase in the ascent velocity of a Taylor bubble enclosed in a vertical tube subjected to horizontal oscillations. In the case of a vertically oscillated tube, \cite{Brannock1996} reported instead experimental evidences of the slowing down of the bubble, while \cite{Madani2009, Madani2012} observed a more nuanced behavior: as the acceleration of the oscillations is gradually increased, the rising of the bubble initially slows down, but then increases at larger accelerations. More recently, \cite{Zhou2024} studied the rising behavior of a Taylor bubble exhibiting volume oscillations imposed either by forcing the liquid column above the bubble to oscillate or by imposing a pulsating pressure field at the top liquid surface. Their numerical simulations evidence that the gas volume oscillations result in the thinning of the lubricating film around the bubble and thus in the decrease of the drainage flow and rising velocity. 

Here, we focus on the transport of Taylor bubbles in sealed tubes filled with a viscous liquid, with an inner radius close to the critical value below which the bubble is stopped in a vertical configuration. We investigate two different strategies to enable bubble motion and tune its velocity, namely rotating the tube around its symmetry axis, and inclining it with respect to gravity (Figure \ref{fig:intro}(c) and (d)). In both cases, we leverage theoretical developments to predict the new threshold for the onset of motion, that depends on the rotational speed and on the tilt angle, respectively. We also provide a prediction for the rising velocity of the bubble as a function of the liquid properties, the tube geometry and the (modified) gravity field. Our theoretical findings are then compared with the outcomes of dedicated experimental campaigns. 

The paper is organized in two parts. In the first part (Section \ref{sec:centrifugation}), we report our investigation on bubble motion in rotating tubes. Section \ref{sec:section_theory_centrifugated} develops the theoretical prediction for the cap profile of the bubble and the matching conditions between this cap and the flat film region, from which we derive the theoretical threshold for the onset of motion and the prediction of the bubble velocity, in terms of the rotational speed. Section \ref{sec:exp_centrifugated} presents the experimental setup and a comparison of the results against the theoretical findings. The second part (Section \ref{sec:inclination}) presents the same structure as the previous one, but investigates the effect of tube inclination, with theoretical predictions and comparison with experimental measurements of bubble transport in tilted tubes.\\

\section{Effect of centrifugation}
\label{sec:centrifugation}
Fluid centrifugation pertains to extensive applications, ranging from the segregation of complex or biological fluids \citep{svedberg1926new} to numerous industrial processes, such as wastewater treatment \citep{Turano2002} or crude oil refining \citep{Gary2007}. In interfacial flows, spinning rods \citep{than1988measurement} and spin-coating \citep{emslie1958flow}, are used to deposit uniform thin films onto diverse substrates such as optical lenses for anti-reflective properties \citep{Krogman2005}, or silicon wafers for organic semiconductors fabrication \citep{Yuan2014}. This method precisely controls the film thickness through the modulation of the angular velocity, essential for achieving high-quality coatings. Additionally, centrifugation can be employed in the generation of surface roughness in curing polymer melts \citep{marthelot2018designing, jambon2021elastic}, where centrifugal instabilities \citep{rietz2017dynamics} are harnessed to facilitate the formation of periodic patterns.

In sealed tubes, the effect of centrifugation on the shape of capillary interfaces has been exploited in spinning drop experiments \citep{vonnegut1942rotating,Rosenthal1962, princen1967measurement,Torza1975}. These experiments can measure very low interfacial tensions \citep{drelich2002measurement}, by rotating a horizontal tube containing a drop of lower-density liquid within a higher-density fluid. For high enough rotation rate, the (transverse) gravity acceleration can be neglected, and the equilibrium shape of the drop results from the balance of the centrifugal force, that tends to elongate the drop along its axis (and thus to thicken the surrounding liquid film), with surface tension, that promotes a spherical shape. 

In this context, \cite{Manning2011} studied the case of a tube of inner radius $R$ partially filled with a liquid of density $\rho$ and surface tension $\gamma$, rotated around its symmetry axis at angular velocity $\omega$ under weightlessness. They derived  a criterion for the occlusion of the tube by a static meniscus spanning the cross-section of the channel, with a contact angle $\phi$, and computed a critical angular velocity $\omega_0$: 
\begin{equation}
\frac{\rho\omega_0^2 R^3}{\gamma}= 32\sin^3\left(\frac{\pi + 2\phi}{6}\right),
\label{eq:manning}
\end{equation}
\noindent such that the tube cannot occlude if $\omega > \omega_0$. In the case of a gas bubble trapped in a capillary, the contact angle $\phi$ is equal to zero: Eq.\eqref{eq:manning} indicates then that under weightlessness, the bubble cannot occlude the channel if $\rho \omega^2 R^3/\gamma > 4$. 

To the best or our knowledge though, the combined effect of axial gravity and transverse centrifugal force on the rising motion of Taylor bubbles in a vertical setting has not been studied yet. Building on the demonstrated ability of centrifugation to elongate light drops or bubbles in tubes, and thereby to thicken their lubricating film, we now study how centrifugation can facilitate the release of Taylor bubbles that are trapped in sealed capillaries due to surface tension.\\

We consider a long bubble of length $L$ immersed in a viscous fluid of dynamic viscosity $\mu$, density $\rho$, and surface tension $\gamma$, both contained in a vertically-oriented circular tube of radius $R \ll L$, sealed at both ends. 
The bubble ascends along the vertical axis at a constant velocity $U_b$ under the influence of gravity.
The tube's radius is assumed to be of the order of the capillary length $\ell_c=\sqrt{\gamma/\rho g}$, where $g$ is the acceleration due to gravity, so that the Reynolds number $\text{Re}=\rho U_b R/\mu$ is sufficiently small to neglect any inertial effects.

Bretherton's solution describing the bubble's ascent at a constant velocity is valid only if the tube radius exceeds a critical value $R_c \approx 0.918 \ell_c$. 
As the tube radius $R$ approaches this critical value, the bubble's ascending speed diminishes, eventually reaching zero. 
This phenomenon can be explained through a simple mass conservation consideration: the sealed tube requires the bubble to displace the liquid below, creating drainage through its peripheral lubricating film. However, for $R < R_c=0.918 \ell_c$, surface tension becomes dominant, causing the bubble to expand and occupy the entire tube cross-section, preventing liquid drainage.

We now examine the scenario where the vertical tube undergoes constant rotation around its symmetry axis with an angular velocity $\omega$. We  can readily anticipate that the centrifugal force will push liquid towards the solid tube wall, thickening the fluid film around the bubble and facilitating its ascent. Therefore,  a steady rising motion of the bubble may be achievable even in tubes with $R< R_c$, provided the angular velocity $\omega$ is sufficiently high. In the subsequent section, we revisit Bretherton's theory \citep{Bretherton1961} to predict the new threshold $R_c(\omega)$ and the steady rising velocity $U_b$ as a function of the rotational speed.

\subsection{Theoretical prediction for the threshold and rising velocity}
\label{sec:section_theory_centrifugated}

Since our focus lies in describing motion near the threshold characterized by a vanishing velocity $U_b$, we preliminary assume a small capillary number $\text{Ca}=\mu U_b/\gamma \ll 1$.
Viscous stresses at the gas-liquid interface thus play a significant role only in regions where the fluid is strongly confined, i.e. where the interface is very close to the solid wall. Consequently, the upper part of the bubble's profile can be divided into two regions, see Figure \ref{fig:intro}(a).
The outer region corresponds to the top of the bubble (cap) where
viscous effects are negligible : the equilibrium is controlled by an interplay between surface tension, gravity, and centrifugal forces. 
Conversely, a thin liquid film resulting from a balance between viscous, surface tension, gravity, and centrifugal forces defines an inner region of small axial curvature.
We now derive the bubble's profiles in these two regions.

\subsection*{The static cap}

In the outer region, the fluid around the cap of the bubble can be considered at rest \citep{Bretherton1961,Lamstaes2017} so that in the cylindrical reference frame co-rotating with the tube and translating with the bubble, the pressure $P$ in the surrounding fluid satisfies: 
\begin{equation}
\label{eq:eq_pressure_static_cap}
\boldsymbol{0}=-\boldsymbol{\nabla} P +\rho \omega^2 r\boldsymbol{e}_r-\rho g \boldsymbol{e}_z. 
\end{equation}

\begin{figure}
\centering
\includegraphics[scale=0.5]{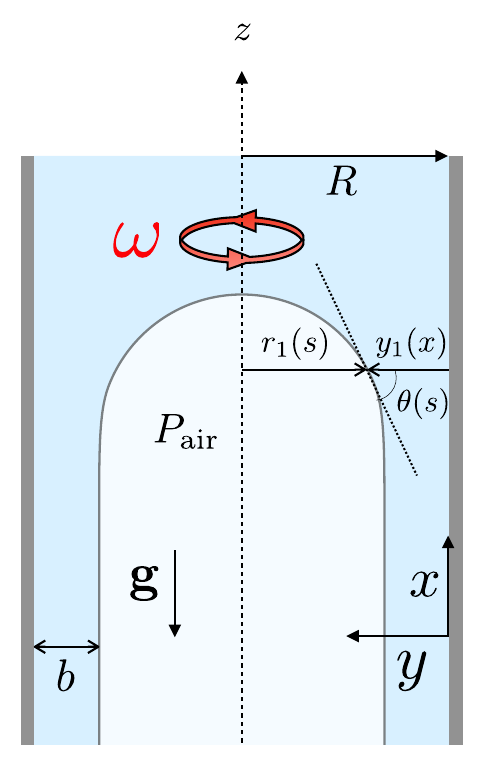}
\caption{Sketch of the bubble in a vertical tube that rotates around its central axis with angular velocity $\omega$. $s$ is the arc-length of the interface measured from the tip of the bubble at $r=0$. In the static cap region, the air-liquid interface is located by the distance $r_1(s)$ to the central axis, and the angle its tangent makes with the horizontal axis is denoted as $\theta(s)$.  In the inner region, where a two-dimensional Cartesian system $(x,y)$ is used, the interface is located instead by its distance from the solid wall $y_1(x)$.}
\label{fig:sketch_bubble}
\end{figure}

By integrating the radial component of Eq. \eqref{eq:eq_pressure_static_cap}, we obtain 
\begin{equation}P(r,z)=\frac{1}{2}\rho  \omega^2\left(r^2-r_1(z)^2\right)+\gamma \kappa +P_{\text{air}},
 \quad \kappa=-\frac{1}{r_1(z)\left(1+r'_1(z)^2\right)^{1/2}}+\frac{r''_1(z)}{\left(1+r'_1(z)^2\right)^{3/2}}\end{equation} where $r_1(z)$ and $\kappa$ denote the location of the air-liquid interface measured from the central axis (oriented upwards) and its curvature, respectively. 

From the axial component of the momentum conservation equation Eq.\eqref{eq:eq_pressure_static_cap}, it follows that:
\begin{equation}
\gamma \kappa-\frac{1}{2} \rho \omega^2r_1(z)^2+\rho g z =\text{cst}.
\end{equation}

By denoting as $s$ the arclength of the interface profile measured from the tip of the bubble and $\theta$ its tangent angle with respect to the horizontal (see Fig. \ref{fig:sketch_bubble}), the static interface profile is given by:
\begin{equation}
-\gamma \left[\frac{d \theta}{ds} +\frac{\sin\theta}{r_1(s)}\right]-\frac{1}{2} \rho \omega^2r_1(s)^2+\rho g z=\text{cst}.
\end{equation}

Differentiating with respect to the curvilinear coordinate $s$ gives: 

\begin{equation}
\label{eq:static_cap_profile}
\gamma \left[\frac{d^2\theta}{ds^2}+\frac{\cos\theta}{r_1(s)}\frac{d\theta}{ds}-\frac{\cos\theta\sin\theta}{r_1(s)^2}\right]=-\rho \omega^2 r_1(s) \cos\theta-\rho g\sin\theta.
\end{equation}

Finally, with the dimensionless variables $\bar{r}_1=r_1/R$ and $\bar{s}=s/R$, Eq. \eqref{eq:static_cap_profile} becomes:
\begin{equation}
\label{eq:final_equation_static_cap_profile}
\frac{d^2\theta}{d\bar{s}^2}+\frac{\cos(\theta)}{\bar{r}_1}\frac{d\theta}{d\bar{s}}-\frac{\cos(\theta)\sin(\theta)}{\bar{r}_1^2}=-\text{Bo}\sin(\theta)-\bar{r}_1\text{Ce}\cos(\theta),
\end{equation}

\noindent where the Bond number $\text{Bo}=\frac{\rho g R^2}{\gamma}=(R/\ell_c)^2$ is introduced as the square of the ratio between the tube radius and the capillary length.
The centrifugal number $\text{Ce}=\frac{\rho \omega^2 R^3}{\gamma}$ can be seen as a rotational Bond number where the centrifugal acceleration $R\omega^2$ plays the role of the gravitational acceleration. 
 
For a given set of parameters $(\text{Bo}, \text{Ce})$, two boundary conditions are required to solve
Eq.\eqref{eq:final_equation_static_cap_profile}. A first condition is provided by the symmetry of the problem, that imposes $\theta(0)=0$ at the top of the static cap. The second boundary condition will be determined upon matching of this static profile with the inner region's one.

\subsection*{The thin film region}

In the inner region where the bubble is surrounded by a thin lubricating film, the film's thickness is extremely small compared to the tube's radius. Following \cite{Bretherton1961}, we thus neglect the azimuthal curvature of the air-liquid interface and consider the thin film region as planar instead of annular.

Under these assumptions, we introduce the two-dimensional, stationary, Cartesian coordinate system $(x,y)$, where $x=z-U_b t$ opposes gravity, and $y=R-r$ represents the distance to the solid wall. In the framework of the lubrication approximation, the viscous flow in the thin film is driven by a pressure gradient resulting from a combination between gravity, capillarity and centrifugal force. 
The axial velocity accordingly writes (See Appendix \ref{app:derivation_velocity_centrigated} for a detailed derivation): 
\begin{equation}
\label{eq:expression_axial_velocity}
u(x,y)= \frac{\gamma}{2 \mu} \left(-y_1''' +\frac{\rho \omega^2 R}{\gamma} y_1' +\frac{\rho g}{\gamma}\right)(y^2-2y_1y)-U_b,
\end{equation}

\noindent where $y_1(x)$ denotes the distance of the air-liquid interface to the solid wall of the tube. In Eq. \eqref{eq:expression_axial_velocity}, the first term of the right-hand-side stems from surface tension effects, the second from the centrifugal force and the third from gravity. Upon integration within the thin film, the volume flux reads: 
\begin{equation}
\label{eq:exp_volume_flux}
Q=-2\pi RU_by_1 -2\pi R \frac{\gamma}{3 \mu} \left(-y_1''' +\frac{\rho \omega^2 R}{\gamma} y_1' +\frac{\rho g}{\gamma}\right)y_1^3.
\end{equation}

This flux must equate the volume of fluid displaced per unit time by the top of the bubble, that is equal to $\pi R^2 U_b$.  Since $y_1/R \ll 1$, the $-2\pi RU_by_1$ term in the expression of the flow rate is a negligible correction. Finally, by imposing flux continuity with the region far away from the tip, where the film thickness can be considered as uniform and equal to a constant $b$, we obtain the following thin film equation:

\begin{equation}
\label{eq:thin_film_equation}
y_1'''=\frac{\rho g}{\gamma}\left(1-\frac{b^3}{y_1^3}\right)+\frac{\rho \omega^2R}{\gamma}y_1'. 
\end{equation}

\noindent Since $b$ is the length scale governing the flow in the inner region, we adimensionalize, as in \cite{Bretherton1961}, with: 
\begin{equation}
y_1=\eta b,\, \,  \, \, x=\zeta b(\rho g b^2/\gamma)^{-1/3},
\end{equation}

\noindent This leads to the ordinary differential equation: 
\begin{equation}
\label{eq: full_non_lin_equation}
\eta'''=\frac{\eta^3-1}{\eta^3}+a\eta'.
\end{equation}

\noindent where $a=\frac{\text{Ce}}{\text{Bo}^{2/3}}\left(\frac{b}{R}\right)^{2/3}$. In Eq. \eqref{eq: full_non_lin_equation}, the left-hand side represents the surface tension term, while on the right-hand side, the first term accounts for gravity, the second for viscous dissipation, and the third for the effect of centrifugation. 

\subsection*{Matching}

We aim at matching
the inner solution, that is described by Eq.\eqref{eq: full_non_lin_equation}, with the static cap solution provided by Eq.\eqref{eq:static_cap_profile}. Given that $b\ll R$, this requires taking the limit $\eta \rightarrow \infty$ in Eq. \eqref{eq: full_non_lin_equation} for the inner solution. In this limit, the equation behaves as: 

\begin{equation}
\label{eq:approx_big_eta}
\eta'''=1+a\eta', 
\end{equation}

\noindent whose general solution reads :

\begin{equation}
\label{en:eq_trans_region_adim_centrifugated}
\eta = c_1 e^{\sqrt{a}\zeta}+c_2e^{-\sqrt{a}\zeta}+c_3-\frac{\zeta}{a}.
\end{equation}

\noindent The values of $c_1$, $c_2$, and $c_3$ can be found through interpolation using the numerical solution of the complete equation Eq.\eqref{eq: full_non_lin_equation}, whose initial conditions are obtained from the uniform film solution. 
Indeed, when $\eta \rightarrow 1$, Eq. \eqref{eq: full_non_lin_equation} becomes 

\begin{equation}
\label{eq:approx_eta_one}
\eta'''=3(\eta-1)+a\eta',
\end{equation}

\noindent for which the only non-oscillating solution is: $\eta_{0}=1 + \mathcal{C}\exp(\mathcal{F}\zeta)$, where $\mathcal{C}$ is an integration constant and: 

\begin{equation}
\mathcal{F}=\frac{\left(\frac{2}{3}\right)^{1/3}a}{\left(27+\sqrt{3}\sqrt{243-4a^3}\right)^{1/3}}+\frac{\left(27+\sqrt{3}\sqrt{243-4a^3}\right)^{1/3}}{2^{1/3}\, 3 ^{2/3}}.
\end{equation}

Since the value of $\mathcal{C}$ can be adjusted by shifting the origin of $\zeta$, we can  set $\mathcal{C}=1$ and 
a large, negative initial value $\zeta_0$ to initialize the integration. This procedure yields initial conditions for the full non-linear equation Eq.\eqref{eq: full_non_lin_equation}, i.e.:
\begin{equation*}
\eta(\zeta_0)=1+\exp(\mathcal{F}\zeta_0), \quad  \eta'(\zeta_0)=\mathcal{F}\exp(\mathcal{F}\zeta_0), \quad \eta''(\zeta_0)=\mathcal{F}^2\exp(\mathcal{F}\zeta_0).
\end{equation*}

\begin{figure}
\centering
\includegraphics[width=\textwidth]{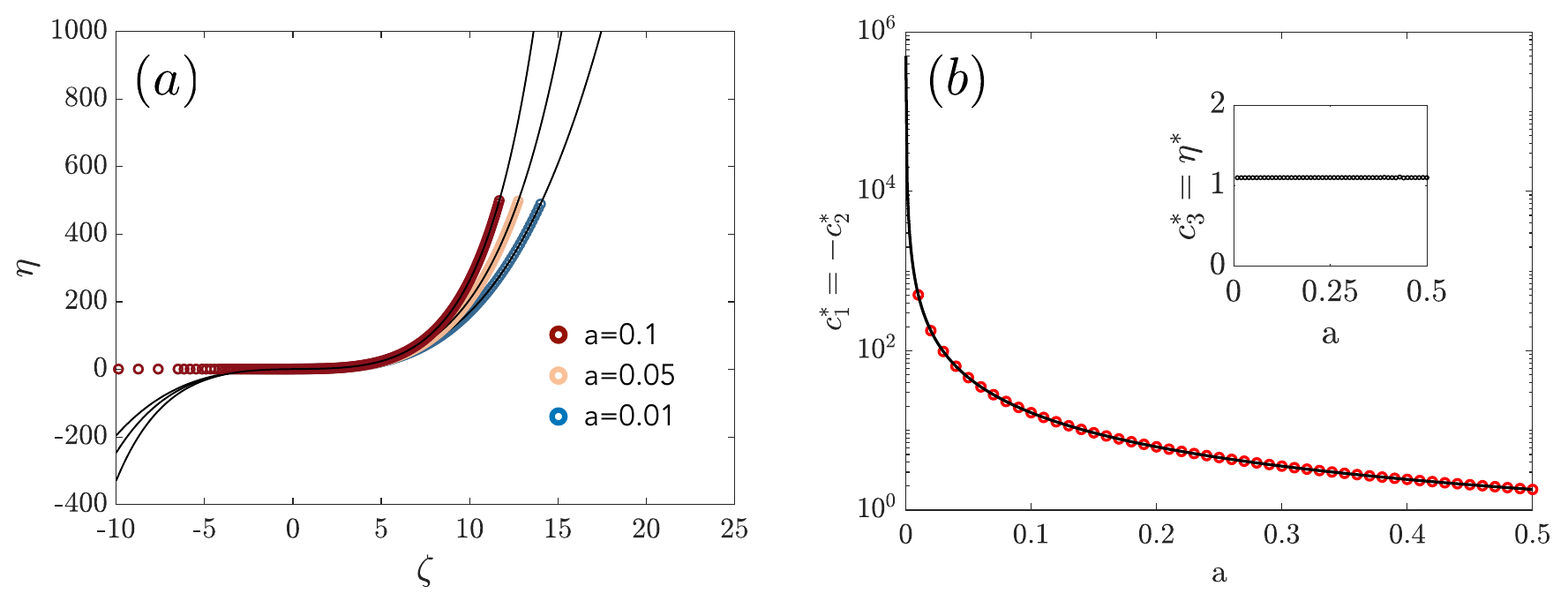}
\caption{(a) The inner region profile $\eta$ as a function of the dimensionless height $\zeta$. The dots represent the solution $\eta$ of the full equation Eq.\eqref{eq: full_non_lin_equation} for various values of $a$, while the black solid lines represent the outer profile of the inner region $\eta = c_1 e^{\sqrt{a}\zeta}+c_2e^{-\sqrt{a}\zeta}+c_3-\frac{\zeta}{a}$, where $c_1(a)$, $c_2(a)$ and $c_3(a)$ are obtained by fitting with the full inner solution, in the $\eta \gg 1$ region. The outer profiles clearly exhibit an inflection point, at a distance referred to as $\zeta^*(a)$. For each value of $a$, the origin is then shifted so that $\eta''(0)=0$. (b) Shifted coefficient $c_1^*=-c_2^*$ as a function of $a$ (red dots). The black solid line corresponds to $ c_1^*=0.500 a^{-3/2}+0.286 a^{-1/2}$. (Inset) Shifted coefficient $c_3^*=\eta(0)\equiv \eta^*$ as a function of $a$. This coefficient does not vary significantly with $a$.}
\label{fig:coefficient_interpolation}
\end{figure}

We solve  Eq.\eqref{eq: full_non_lin_equation}\footnote{Using the built-in \textsc{Matlab} ODE solver \emph{ode45}.} to obtain the inner solution $\eta$ for various values of $a$, and fit the outer profile $\eta = c_1 e^{\sqrt{a}\zeta}+c_2e^{-\sqrt{a}\zeta}+c_3-\frac{\zeta}{a}$ in the region where $\eta \gg 1$. This allows us to retrieve the coefficients $c_1(a)$, $c_2(a)$ and $c_3(a)$. From Figure \ref{fig:coefficient_interpolation}(a), it is evident that the outer profile $\eta$ of the inner solution exhibits an inflection point. We thus translate the origin to the position where $\eta'' =0$, located at the coordinate: 

\begin{equation}
\zeta^*=\frac{1}{2\sqrt{a}}\log\left(-\frac{c_2}{c_1}\right).
\end{equation}

\noindent Using the shifted variable $\chi=\zeta-\zeta^*$, we can now define  $\eta(\chi)=c_1^* e^{\sqrt{a}\chi}+c_2^*e^{-\sqrt{a}\chi}+c_3^*-\frac{\chi}{a}$, where the new coefficients $c_1^*$, $c_2^*$, and $c_3^*$ are expressed as:

\begin{subequations}
\begin{equation}
c_1^*=c_1e^{\sqrt{a}\zeta^*}=c_1\sqrt{-\frac{c_2}{c_1}}=\text{sgn}(c_1)\sqrt{-c_1c_2},
\end{equation}
\begin{equation}
c_2^*=c_2e^{-\sqrt{a}\zeta^*}=c_2\sqrt{-\frac{c_1}{c_2}}=\text{sgn}(c_2)\sqrt{-c_1c_2},
\end{equation}
\begin{equation}
c_3^*=c_3-\frac{\zeta^*}{a}=c_3-\frac{1}{2a^{3/2}}\log\left(-\frac{c_2}{c_1}\right)=\eta(\chi=0).
\end{equation}
\end{subequations}

The new coefficients are well fitted (see Figure \ref{fig:coefficient_interpolation}(b)) by:
\begin{subequations}
\begin{equation}
c_1^*=-c_2^* \approx 0.500 a^{-3/2}+0.286 a^{-1/2},
\end{equation}
\begin{equation}
c_3^*\approx 1.10, \text{ independently of the value of $a$}.
\end{equation}
\end{subequations}

Thus, the distance from the wall at which the outer profile exhibits an inflection point can be evaluated as:
\begin{equation}
y_1(0)=\eta(0)b=(c_1^*+c_2^*+c_3^*)b \approx 1.10b \ll R.
\label{eq:cond_matching_thickness}
\end{equation}

Furthermore, the slope of the profile at the inflection point is given by: 
\begin{equation}
y_1'(0)=\eta'(0)\left(\rho g b^2/\gamma\right)^{1/3}=\left(c_1^* \sqrt{a}-c_2^*\sqrt{a}-\frac{1}{a} \right)\left(\rho g b^2/\gamma\right)^{1/3}=0.572\left(\rho g b^2/\gamma\right)^{1/3}>0.
\label{eq:cond_matching_slope}
\end{equation}

Remarkably, the conditions on film thickness Eq.\eqref{eq:cond_matching_thickness} and slope Eq.\eqref{eq:cond_matching_slope} at the inflection point are the same as described in \cite{Bretherton1961}. Therefore, the centrifugal force alters the inner region solution and the static cap profile, but the matching conditions (and thus the boundary conditions for the static cap solution) remain surprisingly unchanged from those of  \cite{Bretherton1961}.

The above analysis provides the missing information required to solve the static cap profile described by Eq. \eqref{eq:final_equation_static_cap_profile}. 
Specifically, the static profile for a given set of parameters $(\text{Bo}, \text{Ce})$ is obtained through a shooting method, searching for the first derivative $\dot \theta_0$ at the tip of the cap, that is such that the integration of Eq.\eqref{eq:final_equation_static_cap_profile} from initial conditions $(\theta_0=0, \dot \theta_0)$ results in a profile where the inflection point $\ddot r_1(z)=0$ is reached for $r_1(z)=R$. The numerical integration of  Eq.\eqref{eq:final_equation_static_cap_profile} is performed using the \textsc{Matlab} built-in ODE solver \emph{ode23t}, with a spacing along the curvilinear coordinate of 0.01R, while the shooting method is implemented by means of the non-linear \textsc{Matlab} system solver \emph{fsolve}. The resulting slope $r'_1(z)\vert_{r=R}$ at the inflection point is then computed from the generated profile. 

\begin{figure}
\centering
\includegraphics[width=0.8\textwidth]{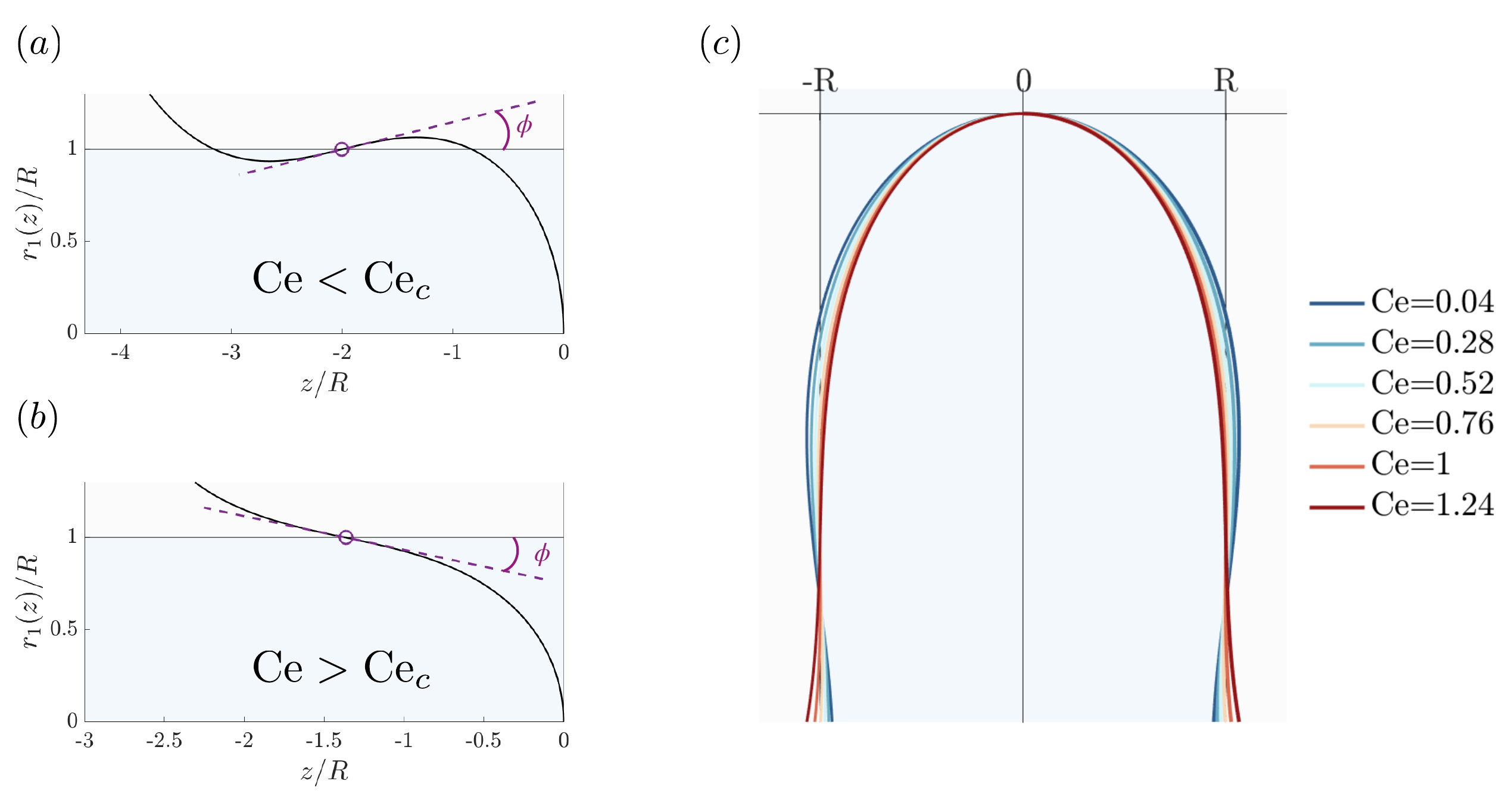}
\caption{Static cap profile for a Bond number $\text{Bo}=0.55$, and a centrifugal number $\text{Ce}$ that is (a) below and (b) above the threshold $\text{Ce}_c(\text{Bo})$. Below the threshold, the slope $r'_1(z)\vert_{r=R}$ is positive at the inflection point, causing the upper profile to escape the fluid domain $r<R$. $\phi$ is the contact angle between the tangent to the static cap profile at the wall and the vertical axis: $\phi=\tan^{-1}\left(-r'_1(z)\vert_{r=R}\right)$ and is thus negative in case (a) and positive in case (b). (c) Evolution of the static cap profile at fixed $\text{Bo}=0.55$, when increasing $\text{Ce}$ from a value below the threshold $\text{Ce}_c\approx 1$ to a value slightly above threshold.}
\label{fig:static_cap_shape}
\end{figure}

For fixed $\text{Bo}<\text{Bo}_{c,0}=(R_c/\ell_c)^2$ and varying $\text{Ce}$ numbers, it appears that some profiles are unphysical: below a critical value $\text{Ce}_c$ that depends on the Bond number $\text{Bo}$, the static cap shape exhibits a positive slope at the inflection point at the solid wall, causing the upper profile to extend beyond the fluid domain $r<R$, as illustrated in Figure \ref{fig:static_cap_shape}, which is not compatible with the matching condition $0<y_1'(0)=-r'_1(z)\vert_{r=R}$. By progressively increasing the centrifugal number $\text{Ce}$, the slope at the inflection point at the wall decreases,  leading to a reduction of the bulge outside the fluid domain, as shown in Figure  \ref{fig:static_cap_shape}(c). Ultimately, for $\text{Ce}>\text{Ce}_c$, the slope becomes negative, causing the entire static cap to reside within the fluid domain, see Figure  \ref{fig:static_cap_shape}(b) and (c). 

\begin{figure}
\centering
\includegraphics[width=\textwidth]{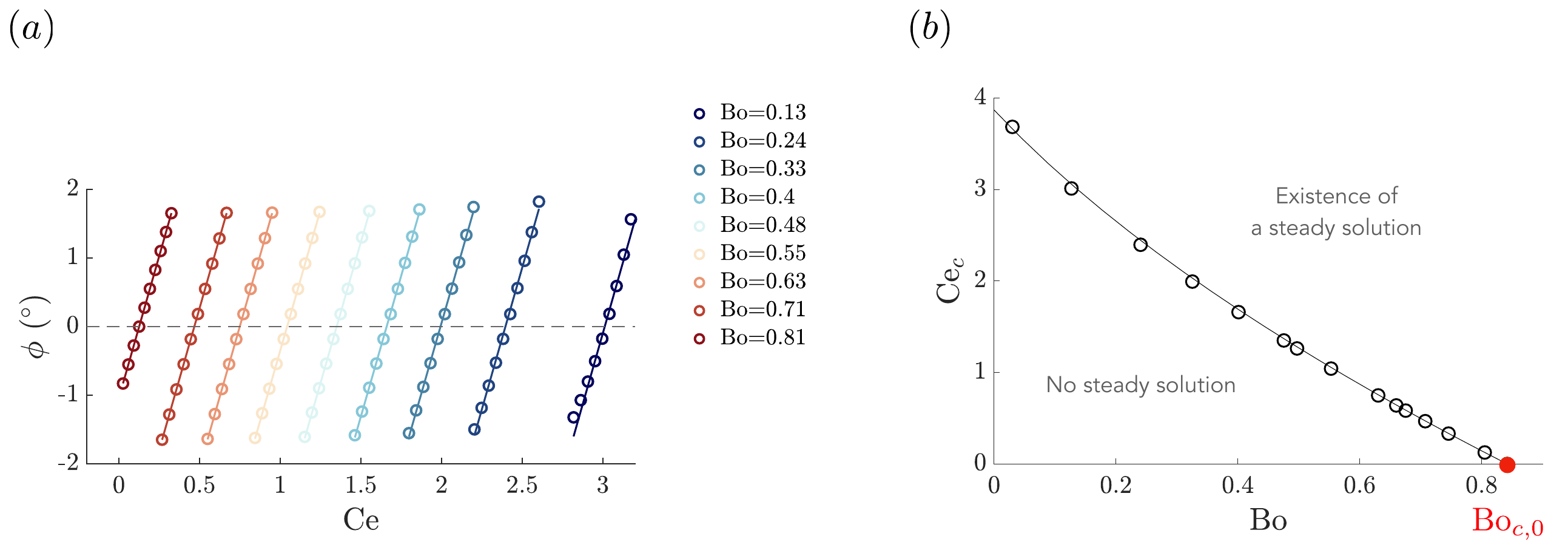}
\caption{(a) Contact angle $\phi$  of the static cap profile between the vertical axis and the tangent to the static cap profile (obtained by integrating Eq. \eqref{eq:final_equation_static_cap_profile} while requiring that the interface reaches the solid wall with an inflection point), as a function of the centrifugal number $\text{Ce}$, for various Bond numbers $\text{Bo}$. The dots are the values computed from the shape of the interface, while the solid lines are the best linear fit. For each $\text{Bo}$, the critical centrifugal number $\text{Ce}_c$ is defined as the value of $\text{Ce}$ for which the best linear fit cancels out. (b) The critical centrifugal number $\text{Ce}_c$ as a function of the Bond number $\text{Bo}$. At a given value of $\text{Bo}$, the matching with the inner region is only possible if $\text{Ce} > \text{Ce}_c(\text{Bo})$. The dots are the values of $\text{Ce}$ that cancel the linear approximation of $\phi(\text{Ce}, \text{Bo})$ for each Bond number, while the dotted back line represents the approximation Eq.\eqref{eq:Ce_crit_vs_Bo}. The red dot with coordinates $(\text{Bo}_{c,0}=0.842, \text{Ce}_c=0)$ locates the threshold in the absence of centrifugation.}
\label{fig:slope_vs_Ce}
\end{figure}

In the following, we denote as $\phi$ the resulting contact angle between the liquid-air interface and the vertical axis, i.e. $\phi=\tan^{-1}\left(-r'_1(z)\vert_{r=R}\right)=\tan^{-1}\left(y_1'(0)\right)$. A closer inspection reveals that within a small range around $\text{Ce}_c$, i.e. for $\vert \text{Ce}-\text{Ce}_c(\text{Bo})\vert <0.2$, $\phi$ varies linearly with $\text{Ce}$, as shown in
Figure \ref{fig:slope_vs_Ce}(a). 
Within this range, the contact angle (in radian) is well approximated by: 
\begin{equation}
\phi(\text{Ce},\text{Bo})\approx 0.144\left(\text{Ce}-\text{Ce}_c(\text{Bo})\right).
\label{eq:slope_vs_Ce}
\end{equation} 

\noindent where the factor 0.144 is independent from $\text{Bo}$ (up to variations less than 0.001 radians). Interestingly, this prediction is consistent with the occlusion criterion derived by \cite{Manning2011} under weightlessness (i.e. $\text{Bo}=0$). Indeed, their equation Eq.\eqref{eq:manning} in the limit of small contact angle becomes: 
\begin{equation}
    \text{Ce}(\phi, \text{Bo=0})\approx 4\left(1+\sqrt{3} \phi\right), 
\end{equation}
\noindent which can be recasted as $\phi(\text{Ce}, \text{Bo}=0)\approx 0.144 \left(\text{Ce}-\text{Ce}_{c,0}\right)$, where $\text{Ce}_{c,0} \equiv \text{Ce}_c(\text{Bo}=0)=4$.

Thus, at fixed $\text{Bo}$, the critical centrifugal number $\text{Ce}_c(\text{Bo})$ for vanishing contact angle (or equivalently for vanishing slope) is retrieved as the value of $\text{Ce}$ at which the best linear fit of $\phi(\text{Ce}, \text{Bo})$ cancels out. Its dependency on the Bond number is depicted in Figure \ref{fig:slope_vs_Ce}(b). Note that we performed a convergence analysis and observed no further variations of $\text{Ce}_c$, within a tolerance of 0.07\%, when increasing 10 times the resolution on the spacing along the curvilinear coordinate used to integrate the static cap profiles.  For $\text{Ce}<\text{Ce}_c$, the geometrical constraint $\phi = \tan^{-1}(y_1(0)) \approx y_1'(0) >0$ cannot be satisfied, so that this value corresponds to the threshold for the onset of motion.

As shown in Figure \ref{fig:slope_vs_Ce}(b), $\text{Ce}_c(\text{Bo})$ is well approximated by:
\begin{equation}
\text{Ce}_c=\frac{0.295-\sqrt{0.295^2-0.080(\text{Bo}_{c,0}-\text{Bo})}}{0.040},
\label{eq:Ce_crit_vs_Bo}
\end{equation}
\noindent where $\text{Bo}_{c,0}=0.842$ is the critical Bond number in the absence of centrifugation ($\text{Ce}=0$). Note that in the limit $\text{Bo} \rightarrow 0$, the approximation Eq.\eqref{eq:Ce_crit_vs_Bo} yields $\text{Ce}_c(\text{Bo}=0)\approx 3.9$, that is close to the threshold $\text{Ce}_{c,0}=4$ computed by \cite{Manning2011} under weightlessness. Conversely, in the limit $(\text{Ce}_c \rightarrow 0$, $\text{Bo}\rightarrow \text{Bo}_{c,0})$, Eq.\eqref{eq:Ce_crit_vs_Bo} simplifies into:
\begin{equation}
\text{Ce}_c\approx \frac{1}{0.295}(\text{Bo}_{c,0}-\text{Bo}).
\label{eq:Ce_crit_vs_Bo_simplified}
\end{equation}
\noindent By injecting Eq. \eqref{eq:Ce_crit_vs_Bo_simplified} into Eq. \eqref{eq:slope_vs_Ce}, we obtain 
\begin{equation}
\phi \approx 0.49\left(\text{Bo}-\left(\text{Bo}_{c,0}-\frac{\text{Ce}}{0.295}\right)\right),
\end{equation} reminiscent of the expression derived by Bretherton for a non-rotating capillary tube ($\phi=0.49\left(\text{Bo}-\text{Bo}_{c,0}\right)$). The similarity of these expressions highlights the role of the centrifugation as a downward shift in the critical Bond number for the onset of motion. \\

In addition, this analysis provides a prediction for the rising velocity of the bubble for $\text{Ce} > \text{Ce}_c(\text{Bo})$. Indeed, 
for $\text{Ce}$ close enough to the threshold $\text{Ce}_c(\text{Bo})$, the slope of the inner solution at the inflection point should verify: 
\begin{equation}
\label{eq:matching_slope}
y_1'(0)=0.572\left(\rho g b^2/\gamma\right)^{1/3}=\tan(\phi)\approx \phi \approx 0.144\left(\frac{\rho \omega^2 (R-1.10.b)^3}{\gamma}-\text{Ce}_c(\text{Bo})\right).
\end{equation}

\noindent Since the volume of fluid displaced per unit time by the tip of the bubble $\pi R^2 U_b$ should be equal to to the volume flux in the uniform film region, we can relate the thickness $b$ to the inner radius $R$ and the velocity $U_b$ through
$\rho g b^3/3\mu U_b=R/2$. Together with Eq.\eqref{eq:matching_slope}, this yields the following expression of $\text{Ca}$ as an implicit function of $\text{Ce}$ and $\text{Bo}$: 
\begin{equation}
\label{theoretical_prediction_velocity_centrifugated}
\text{Ce}-\text{Ce}_{c}=3.78 \text{Ce}\left(\frac{\text{Ca}}{\text{Bo}}\right)^{1/3}+4.35 \text{Bo}^{1/3}\left(\frac{\text{Ca}}{\text{Bo}}\right)^{2/9},
\end{equation}
\noindent where $\text{Ce}_c$ is the function of \text{Bo} described above.  

\subsection{Experiments on centrifugated bubbles}
\label{sec:exp_centrifugated}

In this section, we outline our experimental setup and procedure and compare the results against the above theoretical developments.

\subsection*{Experimental setup and procedure}

Cylindrical borosilicate capillary tubes (Hilgenberg GmbH) are partially filled with silicone oil (Sigma Aldrich, $\rho=964$ kg/m$^3$, $\mu=9.64\times 10^{-2}$ Pa.s, $\gamma=2.09\times 10^{-2}$), leaving an air bubble with a length $L$ greater than 10 times the radius $R$ of the tube. Both ends of the tubes are sealed with epoxy resin. The inner radii of the tubes used in our experimental campaign vary between 0.8 mm and 1.3 mm, corresponding to Bond numbers $\text{Bo}\equiv \rho g R^2/\gamma$ in the range $\left[ 0.29, 0.76\right]$. Note that the uncertainty in the inner diameters of the tubes is of 0.05 mm.

\begin{figure}
\centering
\includegraphics[width=\textwidth]{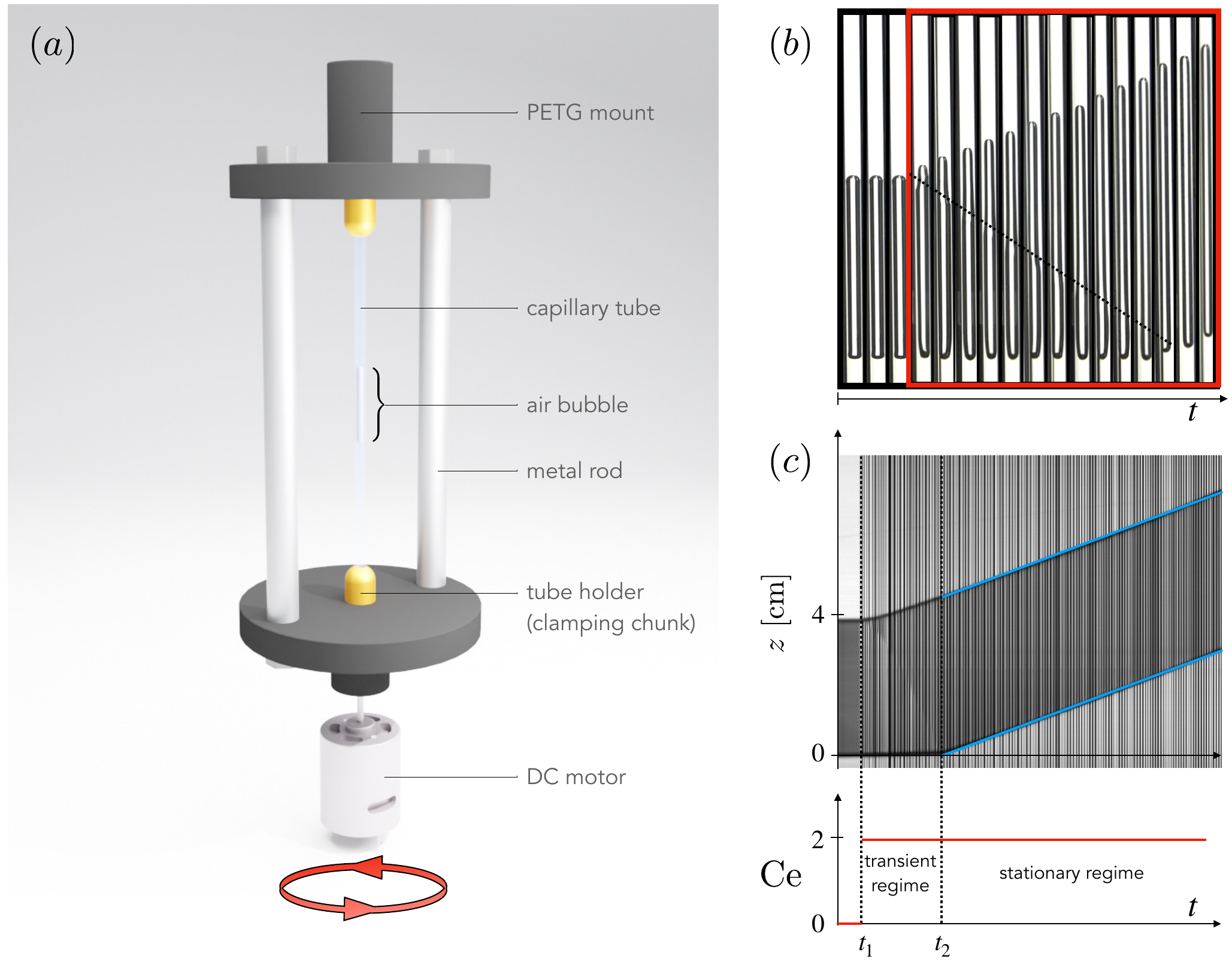}
\caption{Experimental setup and post-processing for rotating bubbles. (a) Tube attachment system. The capillary tube is clamped on both extremities to mounts connected together by two metal rods. The bottom mount is linked to the shaft of a DC voltage-controlled motor that imposes the rotation of the system around its central, vertical axis. (b) Photographs of a long bubble inside a tube filled with silicone oil, at different and equally spaced time steps within the transient regime. In the red frame, the motor has been switched on and the upper cap starts rising while the bottom cap remains still. Along with the resulting bubble elongation, the surrounding liquid film gets progressively thicker from the top to the bottom part of the bubble. The dotted line roughly locates the position of the propagation front. Once the front has reached the lower cap, it starts rising. (c) Intensity profile as a function of time along the tube axis. To produce this image, a column of pixel aligned with the central axis of the tube is extracted from each frame of the movie. The columns are then juxtaposed to each other. The locations of the upper and lower cap as a function of time are easily identified as the two roughy parallel black curves limiting a darker domain that corresponds to the position of the bubble itself. At time $t_1$, the motor is switched on. At $t_2$, the bubble dynamics reaches a stationary state: the upper and lower caps rise at same constant velocity, as highlighted by the parallel blue solid lines that are superimposed on the position of the caps as a function of time. For (b) and (c), $R=1.2$mm and $\text{Ce}=1.97$. The transient duration is approximately equal to $t_2-t_1\approx$ 130s and the capillary number computed from the steady state is $\text{Ca} \approx 3.03 \times 10^{-4}$.}
\label{set_up_centrifugated}
\end{figure}

The tube attachment system, presented in Figure \ref{set_up_centrifugated}(a), consists of two circular mounts made of PETG, rigidly
connected together via
two vertical steel rods, and linked by bearings to a fixed aluminium frame (not represented in the Figure). On each mount, a central, threaded circular mouthpiece accommodates a hollow cylinder whose inner radius matches the outer radius of the capillary tube. The extremities of the tube are then inserted into these cylinders, and securely clamped to the mounts using a clamping chunk. By this means, the tube can be easily replaced by a capillary of a different size with minimal adjustments of the setup. The lower mount is also connected to the shaft of a DC motor that imposes the rotation of the tube attachment system. While the tube, the mounts and the rods rotate collectively, the verticality and stability of the whole setup are ensured by the fixed aluminium frame. 

The motor is voltage-controlled to achieve the desired rotation speed, measured with less than 1\% error using a tachometer. To enhance visualization, a LED panel is positioned behind the tube attachment system.

A Basler camera records the evolution of the bubble in the tube, and the velocities of the upper and lower caps of the bubble are obtained through image post-processing performed via a custom \textsc{Matlab} script. Specifically, a column of pixel aligned with the tube that crosses the upper and bottom profiles of the bubble, is extracted from each frame. These slices are then juxtaposed to each other into an image where the horizontal axis represents time. The displacement 
of the bubble extremities with time
are clearly visible on the resulting image, as shown in Figure \ref{set_up_centrifugated}(c).

Once the motor is switched on, a transient regime occurs
where the upper cap of the bubble rises while the bottom cap remains immobile, resulting in the bubble elongation, reminiscent of spinning bubbles experiments \citep{vonnegut1942rotating}.
This is accompanied by the progressive thickening of the surrounding film that propagates from the top to the lower cap of the bubble, as seen in Figure \ref{set_up_centrifugated}(b).  Once the propagation front reaches the bottom extremity, the lower cap starts its ascent at the same (constant) velocity as the upper cap, see Figure \ref{set_up_centrifugated}(c). The rising regime is assumed to be stationary if the difference between the caps' velocities is less than 5\%. The bubble velocity is computed as the mean velocity between the upper and the bottom cap velocities. 

For a given inner radius $R$ and a fixed rotational speed $\omega$, both $\text{Bo}$ and $\text{Ce}$ are fixed. The capillary number $\text{Ca}=\mu U_b/\gamma$ is then derived from the measurement of the bubble velocity at steady state $U_b$.
For the experiments reported in this Section, we specify that
the Bond number remains smaller than the threshold $\text{Bo}_c=0.842$. Thus, the bubble does not move at all if the rotational speed is zero. 
To avoid excessively long working time for the motor, we did not operate it more than 8 hours consecutively. Considering that with our experimental setup, we cannot precisely detect a motion smaller than 1 mm between the start and the end of an experiment, the smallest capillary number that is experimentally measurable is $\text{Ca}_\text{min}=1.6 \times 10^{-7}$. A value inferior to this limit will be accordingly set equal to zero. The maximal rotational speed achieved by the DC motor is $\omega_\text{max}=400$ rad.s$^{-1}$. For a given tube inner radius, this sets a limit on the maximal centrifugal number $\text{Ce}_\text{max}$ that is experimentally reachable. 

\subsection*{Experimental results and comparison with the theoretical threshold}

\begin{figure}
\centering
\includegraphics[width=\textwidth]{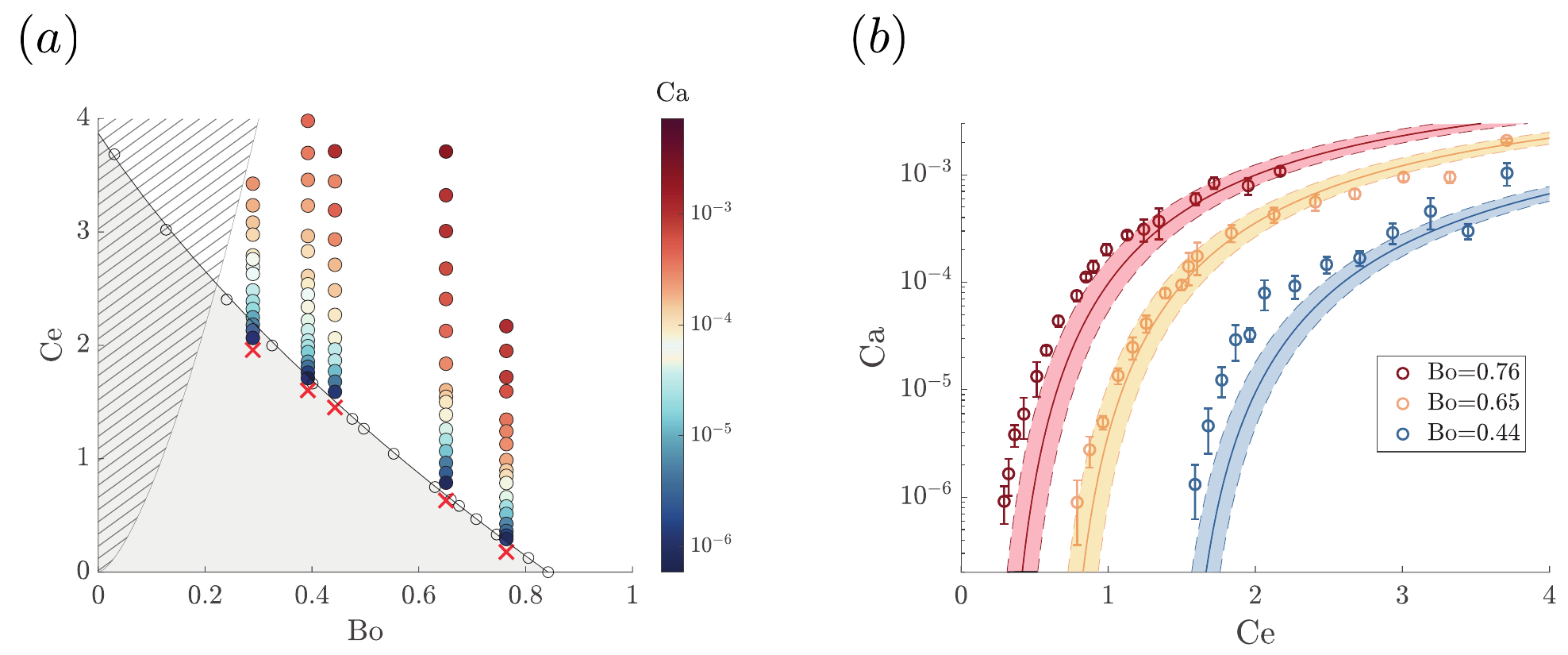}
\caption{(a) Diagram $(\text{Bo}, \text{Ce})$, where each dot corresponds to a measurement of $\text{Ca}>0$ for a given set of parameters $(\text{Bo}, \text{Ce})$. The red crosses indicate the couples $(\text{Bo}, \text{Ce}=\text{Ce}_{c,\text{exp}}(\text{Bo}))$ for which the bubble displacement fell below our detection limit. The black circles indicate the theoretical threshold for the onset of motion and the black solid line represents the approximation Eq.\eqref{eq:Ce_crit_vs_Bo} of $\text{Ce}_c(\text{Bo})$. Below this line,  the gray area indicates the region of parameters where the steady rising of a bubble is not possible according to our theoretical analysis. Finally, the shaded area corresponds to the region of parameters that is not accessible with our setup, due to the constraints on the maximal angular velocity provided by the motor. (b) Capillary number $\text{Ca}$ as a function of the centrifugal number $\text{Ce}$ measured for various Bond numbers. The dots are the experimental points, and for each Bond number $\text{Bo}$, the solid line is the theoretical prediction Eq.\eqref{theoretical_prediction_velocity_centrifugated}, computed using the corresponding experimental value of Bo indicated in the legend. The dotted lines also represent the prediction Eq.\eqref{theoretical_prediction_velocity_centrifugated}, but for $\text{Bo} \pm \Delta \text{Bo}$, where $\Delta$Bo accounts for the $\pm 0.05$ mm uncertainty on the tube inner diameters. The errorbars represent the measurement uncertainty on the bubble velocity.}
\label{exp_results_centrifugated}
\end{figure}

For comparison with the theoretical threshold for the onset of motion, we present our experimental findings in the $(\text{Bo}, \text{Ce})$ diagram featured in Figure \ref{exp_results_centrifugated}(a). Overall, the theoretical prediction is quantitatively consistent with the experimental results, that reveal a rapid decay in the bubble velocity as the centrifugal number $\text{Ce}$ approaches the theoretical threshold $\text{Ce}_c(\text{Bo})$. 
As it was challenging to precisely determine the experimental threshold, we endeavored to establish a narrow range by identifying the highest $\text{Ce}\equiv \text{Ce}_{c,\text{exp}}$ for which the bubble displacement fell below our detection limit. This lower bound is denoted by red crosses on Figure \ref{exp_results_centrifugated}(a), and closely align with the theoretical threshold. However, the prediction is less precise for the smallest values of $\text{Bo}$: the experimental threshold is downward-shifted with respect to the theoretical prediction. 

Figure  \ref{exp_results_centrifugated}(b) reports measurements of the bubble velocity as a function of the rotational speed. The trend is satisfyingly captured by Eq.\eqref{theoretical_prediction_velocity_centrifugated}. We note however that close to the threshold, the measured velocities are in general higher than predicted, consistently with the downward shift of the experimental threshold mentioned above. We believe that this can be ascribed to horizontal vibrations of the tube attachment system observed while operating the motor. As observed by \cite{Kubie2000}, the rising velocity of a Taylor bubble within a vertical tube is indeed larger when the tube is oscillated in the horizontal plane, and increases with the oscillation acceleration. This hypothesis is backed up with the photographs of rotating bubbles along their ascent, that show some asymmetry of the bubble profile with respect to the tube axis, as can be seen for instance in Figure \ref{set_up_centrifugated}(b). This is compatible with the observations of \cite{Kubie2000} under horizontal oscillations: the relative position of the bubble moves periodically from one side of the tube to the other, which thickens the lubricating film on one or the other side of the bubble, resulting in a more efficient drainage and thus in faster bubble ascent. 

Despite these discrepancies, our theoretical analysis seems to provide a good estimation of the threshold for the onset of motion and a satisfying prediction for the general trend of the rising velocity as a function of the rotational speed.
 \\

In summary, rotation reduces the critical tube radius for the onset of motion and facilitates bubble ascent. 
From a theoretical point of view, the most appreciable effect of centrifugation is the modification of the static cap profile, while geometrical constraints stemming from the matching with the thin film region appear remarkably unchanged from the classical case without rotation.
At the same time, experiments demonstrate the thickening of the thin film surrounding the elongated part of the bubble, for increasing rotational speed. This thickening is caused by the centrifugal acceleration which induces a radial, "gyrostatic" pressure gradient that pushes liquid towards the solid wall.
We can thus interpret centrifugation as a mean to tune the thickness, and thus the flow rate within the gap between the tube wall and the bubble, resulting in the lowering of the critical Bond number for the onset of motion.

An alternative and simple strategy to modify the hydrostatic pressure gradient is to tilt the tube with respect to gravity, whose effect is investigated in the next Section.

\section{Effect of tilt }
\label{sec:inclination}

The influence of inclination angle on the mobility of elongated bubbles was first observed by \cite{White1962}, who 
pointed out the necessity of careful positioning of the pipes for precise measurement of the rising velocity. Since then, many studies have been dedicated to the motion of long bubbles in inclined pipes (\cite{Zukoski1966}, \cite{Maneri1974}, \cite{Bendiksen1984}, \cite{Weber1986}, \cite{Couet1987}, \cite{Shosho2001}, \cite{Boucher2023} among others). 
All studies reported a non-monotonous dependency of the rising velocity on the tilt angle: starting from a horizontal position, the velocity of elongated bubbles increases with the inclination of the pipe, reaching a maximum value. Subsequently, the velocity decreases until the vertical position is attained. These observations are reminiscent of the so-called Boycott effect \citep{Boycott1920,Acrivos1979} in the case of settling suspensions in sealed tubes, as seminally observed by \cite{Boycott1920} with blood corpuscules sedimenting in serum, that demonstrated a several-fold increase in their sedimentation rate when the tube was inclined. 

Most of these analyses
are interested in the inertial regime, with large Bond numbers, and only scarce studies were dedicated to the regime close to the onset of motion, that is dominated by surface tension (low Bond number).
\cite{Zukoski1966} conducted an extensive series of experiments focusing on the velocity of elongated bubbles in tubes within a large range of Bond numbers, delving into the impact of liquid viscosity and surface tension on bubble velocity. For low Bond number ($\text{Bo}=0.870$), elongated bubbles exhibited no detectable movement in horizontal or vertical positions but could rise in inclined tubes with angles ranging from $20\degree$ to $80 \degree$ with the horizontal, with a maximum velocity reached about $50 \degree$. This observation suggests that tilting the tube with respect to gravity may enable the motion of long bubbles that are stuck in a vertical configuration owing to surface tension. 

In a similar context, \cite{Collicott2014} studied the stability of a liquid mass in a tube above a capillary interface spanning the cross-section of the channel, for various contact angles and tube inclinations with respect to gravity. At fixed contact angle, they computed the critical Bond number as the threshold above which the \textsc{Surface Evolver} simulations do not converge to a solution of finite axial extent, thereby identifying the critical Bond number as a stability threshold for the capillary interface. Within this approach, the critical Bond number for a $0\degree$- contact angle should correspond to the stability threshold of a long static bubble, expanding over the entire cross-section of the tube. However, difficulties of modelling perfectly wetting conditions prevented the authors to compute the critical Bond number as a function of inclination in this case. 

To the best of our knowledge, there is currently no predictive analysis of the mobility enhancement of Taylor bubbles due to tilted gravity in very narrow capillaries. In this study, we investigate how the direction of gravity affects the mobility of long bubbles in the low $\text{Bo}$ regime, focusing on the angle-dependent threshold for the initiation of motion.

\subsection{Theoretical prediction for the threshold and rising velocity}

\subsection*{The three-dimensional static cap}

We first introduce the equilibrium equation for the static three-dimensional shape
of the upper cap of the bubble. We define a Cartesian coordinate system $(x,y,z)$ where $z$ is the direction aligned with the central axis of the tube. The gravity vector reads $\mathbf{g} = \left(g\cos(\alpha), 0, -g\sin(\alpha)\right)$, where $\alpha$ is the tilt angle of the tube ($\alpha=90\degree$ corresponds to a vertical tube), see Figure \ref{fig:config_validation_inclined}(a). 

The evaluation of the static interface profiles of the upper cap is based on the two-dimensional Young-Laplace equation, where length scales are non-dimensionalized with the tube radius \citep{Manning2011,Rascon2017}:
\begin{equation}
\nabla\cdot\left(\frac{\nabla \bar{h}}{\sqrt{1+(\nabla \bar{h})^2}}\right)=\text{Bo}\left(\cos(\alpha)\bar{x}-\bar{h}\sin(\alpha)\right),
\label{eq:YL_equation}
\end{equation}

\noindent where $\bar{h}(\bar{x},\bar{y})$ denotes the height of the static cap; the quantity on the left hand side is the curvature $\kappa$ of the liquid-gas interface while the term on the right hand side corresponds to the hydrostatic contribution. Eq.\eqref{eq:YL_equation} is complemented with the boundary condition at the solid wall: $ \frac{\nabla h}{\sqrt{1+(\nabla h)^2}}.\mathbf{n}=-\cos(\phi)$, where $\mathbf{n}$ is the outwards-oriented vector normal to the tube, and the contact angle $\phi$ is defined similarly as in the first part of this study, as the angle between $\mathbf{e}_z$ and the tangent at the wall to the intersection of the liquid-gas interface with the plane $(\mathbf{n}, \mathbf{e}_z)$. We thus require the interface to reach the solid wall with a specified slope, that is assumed to be the same for all directions $\mathbf{n}$. Note that this differs from the first part of this study, where we imposed a vanishing radial curvature at the wall and computed the contact angle $\phi$ a posteriori. Here, the contact angle is specified as a boundary condition, with no requirement on the curvature. We acknowledge that requiring the gas-liquid interface to reach the wall for all directions $\mathbf{n}$ is somehow counterintuitive, given that for small tilt angles $\alpha$ (i.e. for a strongly inclined tube with respect to gravity), we expect the film surrounding the bubble to be thicker in the direction $x>0$, and the liquid-gas interface to be relatively far from the solid wall in this region. However, we focus here on the vicinity of the threshold for the onset of motion, where surface tension is dominant and causes the bubble to expand in the entire fluid domain $\sqrt{x^2+y^2}=r<R$ in all directions, as experimentally observed (see for instance Figure \ref{set_up_inclined}(d)). The derivation of Eq.\eqref{eq:YL_equation} can be found in Appendix \ref{app:young_laplace_equation}.

\subsection*{The thin film region and matching}

\begin{figure}
\centering
\includegraphics[width=\textwidth]{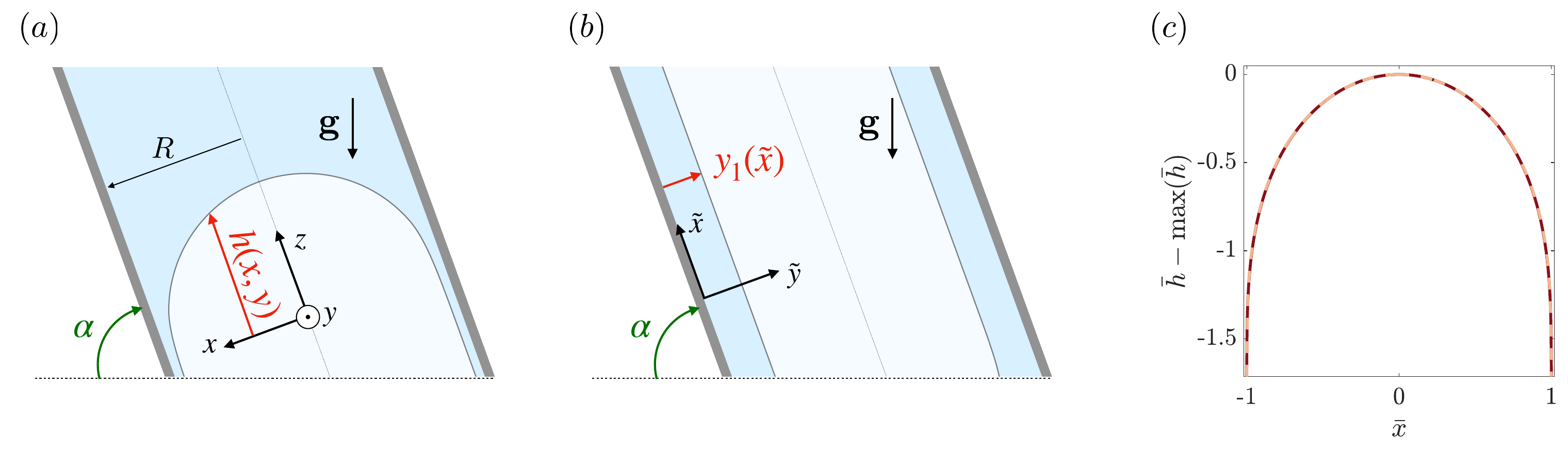}
\caption{(a) Sketch of the static cap in a tube tilted with angle $\alpha$ with respect to the horizontal plane. A Cartesian coordinate system $(x,y,z)$ is used, where $z$ is the direction aligned with the central axis of the tube. The height of the liquid-air interface is denoted as $h(x,y)$ (b) Sketch of the thin film region. A two-dimensional Cartesian coordinate system $(\tilde{x},\tilde{y})$ is used, where $\tilde{x}$ is the direction aligned with the central axis of the tube. The distance of the liquid-air interface from the solid wall is denoted by $y_1(\tilde{x})$. (c) Comparison between the solution of Eq.\eqref{eq:YL_equation} (red solid line) and the axisymmetric solution of Eq.\eqref{eq:static_cap_profile} with no rotation ($\omega=0$) (pink dotted line), for a tilt angle $\alpha=90\degree$ (vertical tube), a contact angle $\phi=0.50\degree$ and a Bond number $\text{Bo}=0.86$.}
\label{fig:config_validation_inclined}
\end{figure}

Here, we opt for a simplified description of the inner region, where we neglect azimuthal variations of curvature and film thickness. That assumption allows us to use as before a two-dimensional, stationary Cartesian reference frame $(\tilde{x}=z-U_bt,\tilde{y})$, where $\mathbf{e}_{\tilde{x}}$ is aligned with the tube central axis and points upwards (such that $\mathbf{e}_{\tilde{x}}\cdot\mathbf{g}=-g\sin(\alpha)$), and $\mathbf{e}_{\tilde{y}}$ is the inward vector normal to the inner solid wall, such that $\mathbf{e}_{\tilde{y}}\cdot\mathbf{g}=-g\cos(\alpha)$, see Figure \ref{fig:config_validation_inclined}(b). 

Within the lubrication framework, the viscous flow in the thin film around the bubble is driven by Laplace and hydrostatic pressure gradients. The axial velocity accordingly writes (see Appendix \ref{app:derivation_velocity_tilted} for a detailed derivation): 
\begin{equation}
u(\tilde{x},\tilde{y})=\frac{\gamma}{2\mu}\left[-y_1'''+\frac{\rho g \cos(\alpha)}{\gamma}y_1'+\frac{\rho g \sin(\alpha)}{\gamma}\right]\left(\tilde{y}^2-2y_1\tilde{y}\right)-U_b,
\end{equation}
\noindent where $y_1$ denotes the distance of the air-liquid interface to the solid wall of the tube. Owing to volume conservation, the flow rate must verify:
\begin{align}
Q&=-2\pi RU_b y_1-2\pi R\frac{\gamma}{3\mu}\left(-y_1'''+\frac{\rho g\cos(\alpha)}{\gamma}y_1'+\frac{\rho g\sin(\alpha)}{\gamma}\right)y_1^3\\
\label{eq:flux_continuity_1}
&=-2\pi RU_b b-2\pi R \frac{\rho g \sin(\alpha)}{3\mu}b^3\\
\label{eq:flux_continuity_2}
&=-\pi R^2U_b.
\end{align}
\noindent where the first and second equality correspond to the volume flux in the inner region and in the uniformly thin film region, respectively, while the third equality describes the volume of fluid displaced per unit time by the top of the bubble.  Since $y_1/R\ll 1$ and  $b/R \ll 1$, the $-2\pi RU_by_1$ and $-2\pi RU_b b$ terms are negligible corrections, which leads to the following thin film equation: 
\begin{equation}
y_1'''=\frac{\rho g \sin(\alpha)}{\gamma}\left(1-\frac{b^3}{y_1^3}\right)+\frac{\rho g \cos(\alpha)}{\gamma}y_1'.
\end{equation}

We adimensionalize with: 
\begin{equation*}
y_1=\eta b, \quad \tilde{x}=\zeta b(\rho g b^2\sin(\alpha)/\gamma)^{-1/3},
\end{equation*}
\noindent which leads to the ordinary differential equation: 
\begin{equation}
\eta'''=\frac{\eta^3-1}{\eta^3}+a\eta',
\end{equation}
\noindent where $a=\cos(\alpha)\sin(\alpha)^{-2/3}\text{Bo}^{1/3}\left(\frac{b}{R}\right)^{2/3}$. 

Upon introduction of the parameter $a$, this equation is exactly the same as Eq.\eqref{eq: full_non_lin_equation} describing the inner region in a centrifugated tube, that has been solved in Section \ref{sec:section_theory_centrifugated}. Thus, shifting the origin to the position where $\eta''=0$, the distance from the wall at which the inner solution exhibits an inflection point is:
\begin{equation}
y_1(0)=\eta(0)b=1.10 b \ll R, 
\end{equation}
\noindent and the slope of the inner region profile at the inflection point is given by:
\begin{equation}
y_1'(0)=\eta(0)\left(\rho g b^2\sin(\alpha)/\gamma\right)^{1/3}=0.572 \, \text{Bo}^{1/3}\sin(\alpha)^{1/3}\left(\frac{b}{R}\right)^{2/3} >0.
\label{eq:matching_slope_tilted}
\end{equation}

For the matching of the thin film region with a two-dimensional cap, we would need to determine the value of (positive) contact angle which leads to zero curvature at the wall. For the matching with the previously introduced three-dimensional shape of the static cap, we extend this analysis by searching for the contact angle that gives rise to zero radial curvature in \textit{at least one point} of the matching boundary. Note that although we imposed as a boundary condition a constant contact angle $\phi$ at the interface when reaching the wall, the height of the interface at the wall, and so the curvature, vary along the azimuthal direction. Since the contact angle at the point of vanishing curvature should be positive according to the matching condition Eq.\eqref{eq:matching_slope_tilted}, we identify the critical Bond number as the $\text{Bo}$ value for which the contact angle is zero, and the radial curvature at the wall vanishes in at least one point. For smaller Bond numbers, the geometrical constraint on the slope cannot be satisfied in at least one point of the domain. 

\begin{figure}
\centering
\includegraphics[width=\textwidth]{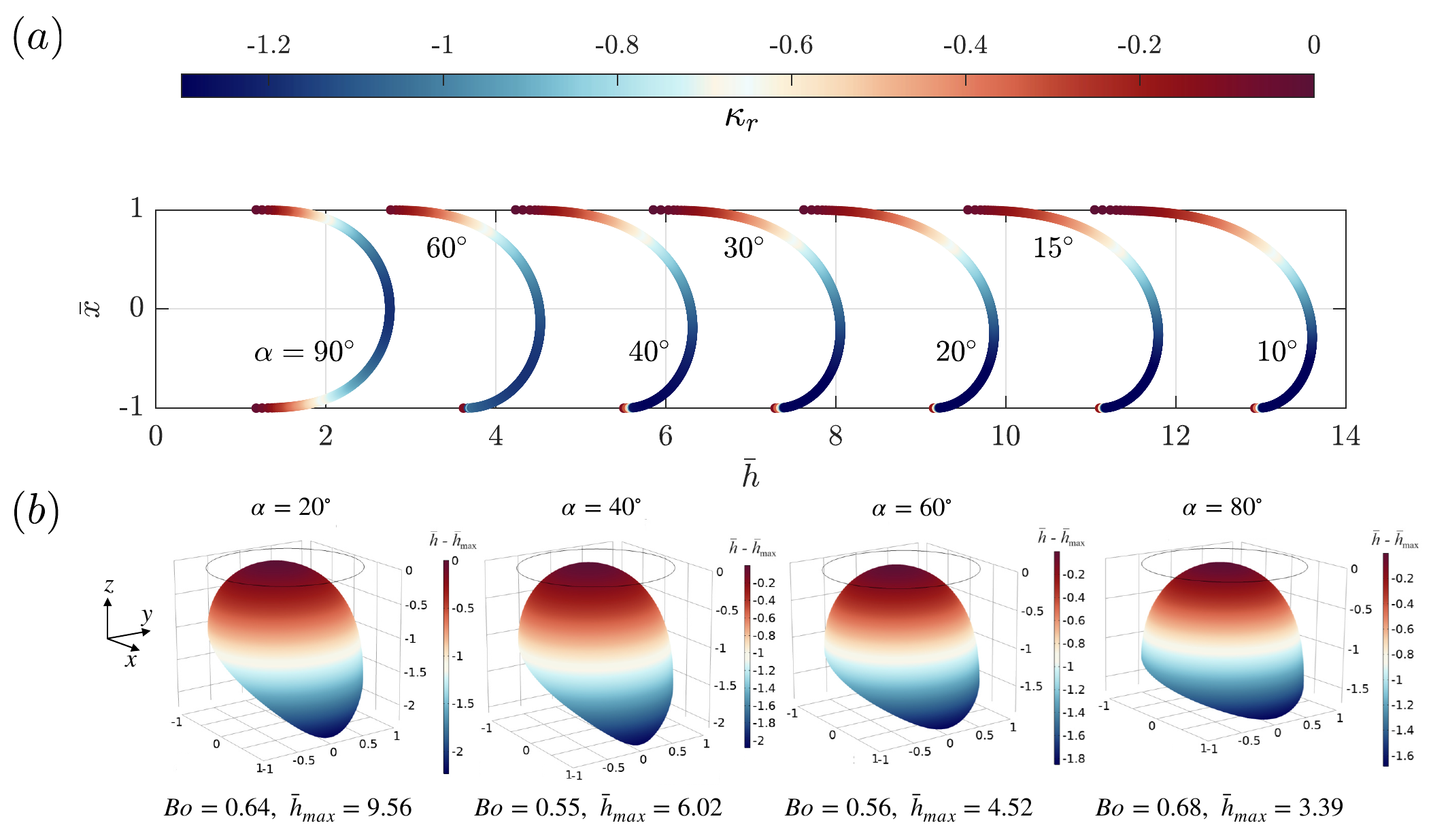}
\caption{(a) Static cap profiles computed as solutions of Eq.\eqref{eq:YL_equation} for various tilt angles $\alpha$, with $\phi=0.5\degree$ and $\text{Bo}\gtrapprox \text{Bo}_c(\alpha)$. The colorbar represents the radial curvature $\kappa_r$ computed along the height profiles in the plane $y=0$. The heights of the profiles for various $\alpha$  have been translated for visualisation purposes. (b) Three-dimensional static cap shape close to critical conditions $\phi=0.5\degree$, for various tilt angles. The colorbar represents the profile height  $\bar{h}-\bar{h}_\text{max}$. }
\label{fig:critical_static_cap_profile_vs_alpha}
\end{figure}

Eq.\eqref{eq:YL_equation} together with its boundary conditions, is implemented in the Finite-Elements solver \textsc{Comsol} Multiphysics. We exploit fourth-order Lagrangian shape functions, solving for the height $h$ and the mean curvature $\kappa$ in a grid composed of quadrangolar elements, with 10 boundary layers of 1.3 stretching factor to properly capture the curvature at the boundary. For each tilt angle $\alpha$, solutions are obtained for different values of the contact angle $\phi$ and Bond number $\text{Bo}$ using the built-in Newton algorithm, initialized with the zero solution. We then perform a continuation study by gradually decreasing the angle $\phi$ from 90$\degree$. Note that the boundary condition $\phi=0\degree$ cannot be imposed in this framework, as it implies infinite directional derivatives for the thickness. We thus study solutions in the close vicinity of $\phi=0\degree$ and extrapolate the retrieved behavior for vanishing contact angles; however, this limitation will not significantly affect the evaluation of the threshold for the bubble rise. 
For fixed Bond number, a convergence analysis from a characteristic size of 0.05$R$ to 0.01$R$ (i.e. from 3200 to 33012 elements) showed variations of $\sim 10^{-4}$ rad in the value of the contact angle resulting in a zero radial curvature. 
The numerical code for  $\alpha=90\degree$ (i.e. for a vertical tube), $\phi=0.5\degree$ and $\text{Bo}=0.86$, is compared against the axisymmetric solution of Eq.\eqref{eq:static_cap_profile} with {no rotation} ($\omega=0$). The result of the comparison is reported in Figure \ref{fig:config_validation_inclined}(c) and a good agreement is observed.

Figure \ref{fig:critical_static_cap_profile_vs_alpha} shows the static cap of the bubble for different inclination angles and Bond numbers, for same contact angle $\phi=0.5\degree$.
For $\alpha<90\degree$, the static cap is not axisymmetric: as the tilt angle increases, the apex of the cap moves toward negative $x$. Conversely, an elongated region (tongue) develops in the vicinity of the $x$ axis, in the direction $x>0$ (i.e. in the direction of positive gravitational acceleration)  and becomes longer as the tilt angle increases. The elongated region presents abnormal values of mean curvature with respect to the rest of the cap. The highest (negative) curvature is observed to be localized at the tip point of this tongue.

To obtain the critical conditions, we fix the tilt angle and the Bond number and progressively decrease the contact angle $\phi$. A preliminary analysis showed that the highest radial curvature is obtained at the tip point of the tongue (of coordinates $(\bar{x}=1,\bar{y}=0)$), in agreement with the above observations, and increases with decreasing contact angle. For each contact angle, we thus compute the radial curvature $\kappa_r=\frac{\partial^2 \bar{h}}{\partial \bar{x}^2}/{\left(1+\left(\frac{\partial \bar{h}}{\partial \bar{x}}\right)^2\right)^{3/2}}$ at the extremity of the tongue. Note that $\frac{\partial \bar{h}}{\partial \bar{y}} (\bar{x},\bar{y}=0)=0$ because of symmetry. 
The contact angle is then decreased until $\kappa_r$ vanishes. This limit value\footnote{The limit value is obtained through linear interpolation, when a change of sign is detected,  of the values of curvature between two successive values of $\phi$, with a step of $9 \times 10^{-5}$ rad.} of contact angle is denoted $\phi_\text{lim}(\alpha, \text{Bo})$. For the set of parameters $(\text{Bo}, \alpha, \phi=\phi_\text{lim}(\alpha, \text{Bo}))$, the liquid-air interface exhibits then an inflection point \textit{at the wall}, and its tangent plane makes an angle $\phi_\text{lim}(\alpha, \text{Bo})$ with the $z$-direction.

\begin{figure}
\centering
\includegraphics[width=\textwidth]{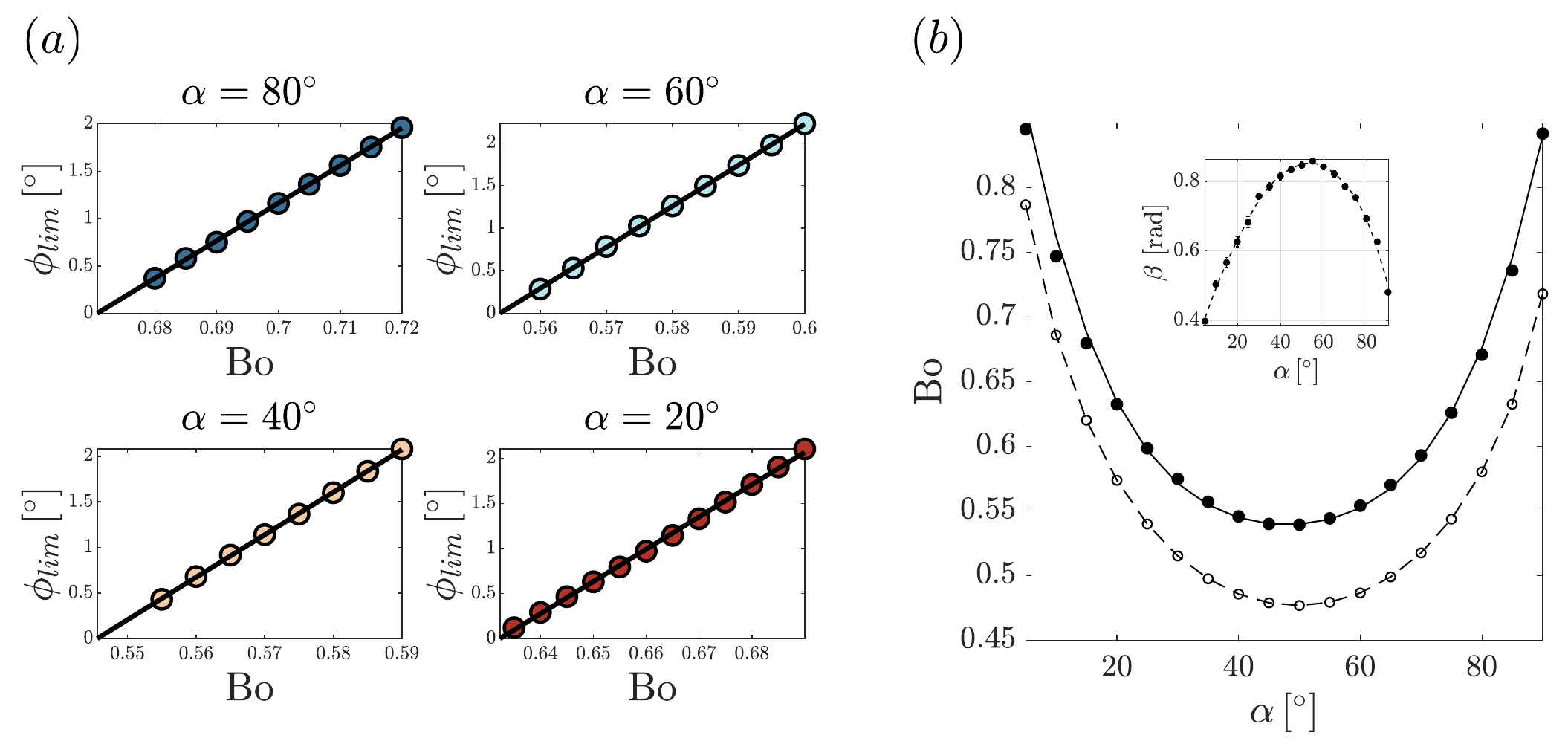}
\caption{(a) Contact angle value $\phi_\text{lim}$  for which the radial curvature of the liquid-air interface vanishes at one point at the wall, as a function of $Bo$ for various values of $\alpha$ (colored dots). For each panel, the black solid line is the linear fit $\phi_\text{lim}(\alpha, \text{Bo})=\beta(\alpha)\left[\text{Bo}-\text{Bo}_c(\alpha)\right]$ performed in order to retrieve the threshold for the onset of motion $\text{Bo}_c(\alpha)$, that corresponds to $\phi_\text{lim}=0$. (b) Threshold $\text{Bo}_c$ as a function of the tilt angle $\alpha$ (black dots). The maximum extrapolation error of the order of 0.001  and is smaller than the marker size. The black solid line represents the polynomial approximation Eq.\eqref{eq:poly_approx_BoCr_vs_alpha}. The black circles represent instead the threshold $\text{Bo}^{2D}_c(\alpha)$ retrieved from the matching of the thin film with a two-dimensional static cap profile. The black dotted line is a guide for the eyes. (Insert) Coefficient $\beta$ as a function of the tilt angle $\alpha$. The errorbars represent the 95\% confidence interval. }
\label{fig:phi_vs_alpha_Bo_vs_alpha_theorique}
\end{figure}

Repeating the same procedure varying the Bond number while fixing the tilt angle $\alpha$, we can retrieve $\phi_\text{lim}$ as a function of $\text{Bo}$. In the range $\phi_\text{lim} \in \left[0.5\degree, 2 \degree\right]$, $\phi_\text{lim}(\alpha, \text{Bo})$ varies linearly with the Bond number $\text{Bo}$, as shown in Figure \ref{fig:phi_vs_alpha_Bo_vs_alpha_theorique}(a). For each tilt angle $\alpha$ we interpret the Bond number value at which $\phi_\text{lim}(\alpha, \text{Bo}_c)=0$, as the threshold $\text{Bo}_c(\alpha)$ for the onset of motion. We retrieve this value by performing for each tilt angle $\alpha$, a linear fit of  $\phi_\text{lim}(\alpha, \text{Bo})$ for $\phi_\text{lim}$ varying between 0.5$\degree$ and 2$\degree$\footnote{The fit is performed by considering at least eight points within the declared range. We verified that the threshold and slope do not vary appreciably by decreasing the number of points while keeping a constant distance between the remaining points.}, and by extrapolating the value $\text{Bo}_c(\alpha)$ that corresponds to $\phi_\text{lim}=0\degree$. The slope of the fit is also a function of $\alpha$, so that overall, $\phi_\text{lim}(\alpha,\text{Bo})$ is approximated by: 
\begin{equation}
\phi_\text{lim}(\alpha, \text{Bo})=\beta(\alpha)\left(\text{Bo}-\text{Bo}_c(\alpha)\right).
\end{equation}

\noindent The result of this procedure is displayed on Figure \ref{fig:phi_vs_alpha_Bo_vs_alpha_theorique}. Our study clearly indicates that the threshold for the onset of motion is lowered by tilting the tube, with a minimum that is reached for a tilt angle of $45\degree<\alpha_\text{opt}<50\degree$. Overall, the critical Bond number as a function of the tilt angle is well approximated by:
\begin{equation}
   \text{Bo}_{c}(\alpha)\approx 0.54\left[1 + 0.5\left(\frac{\pi}{180}\right)^2(\alpha -48\degree)^2 + \left(\frac{\pi}{180}\right)^4(\alpha-48\degree)^4\right],   
   \label{eq:poly_approx_BoCr_vs_alpha}
\end{equation}
\noindent as shown in Figure \ref{fig:phi_vs_alpha_Bo_vs_alpha_theorique}(b). Note that in the vertical case, we retrieved values for the critical Bond number $\text{Bo}_{c}(\alpha=90\degree)$ and for the slope $\beta(\alpha=90\degree)$, that match Bretherton's values within 0.1\% and 1\% of relative error, respectively,  thus validating further the procedure.
\\

We now aim at providing a prediction for the ascent velocity. The matching of the two-dimensional thin film region with the static cap shape at the inflection point amounts to enforce:
\begin{equation}
0.572 \, \text{Bo}^{1/3}\sin(\alpha)^{1/3}\left(\frac{b}{R}\right)^{2/3}=\phi_\text{lim}(\alpha, \text{Bo})=\beta(\alpha)\left[\frac{\rho g}{\gamma}(R-1.10b)^2-\text{Bo}_c(\alpha)\right].
\end{equation}

From the volume conservation constraint Eq.\eqref{eq:flux_continuity_1}-\eqref{eq:flux_continuity_2}, $(b/R)=\left(\frac{3\text{Ca}}{2\text{Bo}\sin(\alpha)}\right)^{1/3}$, which finally yields the following implicit function for the bubble velocity: 

\begin{equation}
\label{eq:prediction_velocity_inclined}
\text{Bo}-\text{Bo}_c(\alpha)=2.52\, \text{Bo}^{2/3}\text{Ca}^{1/3}\sin(\alpha)^{-1/3}+\frac{0.63}{\beta(\alpha)}\text{Bo}^{1/9}\text{Ca}^{2/9}\sin(\alpha)^{1/9}.
\end{equation}

\subsection{Experiments on tilted bubbles}

\subsection*{Experimental setup and procedure}

The same silicone oil used in Section \ref{sec:exp_centrifugated} is employed to partially fill capillary tubes, that are then sealed on both ends, trapping a long air bubble inside. The inner radii of the tubes vary between 1.08mm and 1.96mm, corresponding to Bond numbers in the range $\text{Bo} \in \left[0.53, 1.73\right]$. As in the previous Section, the uncertainty in the tubes inner diameters is of 0.05 mm.

\begin{figure}
\centering
\includegraphics[width=0.9\textwidth]{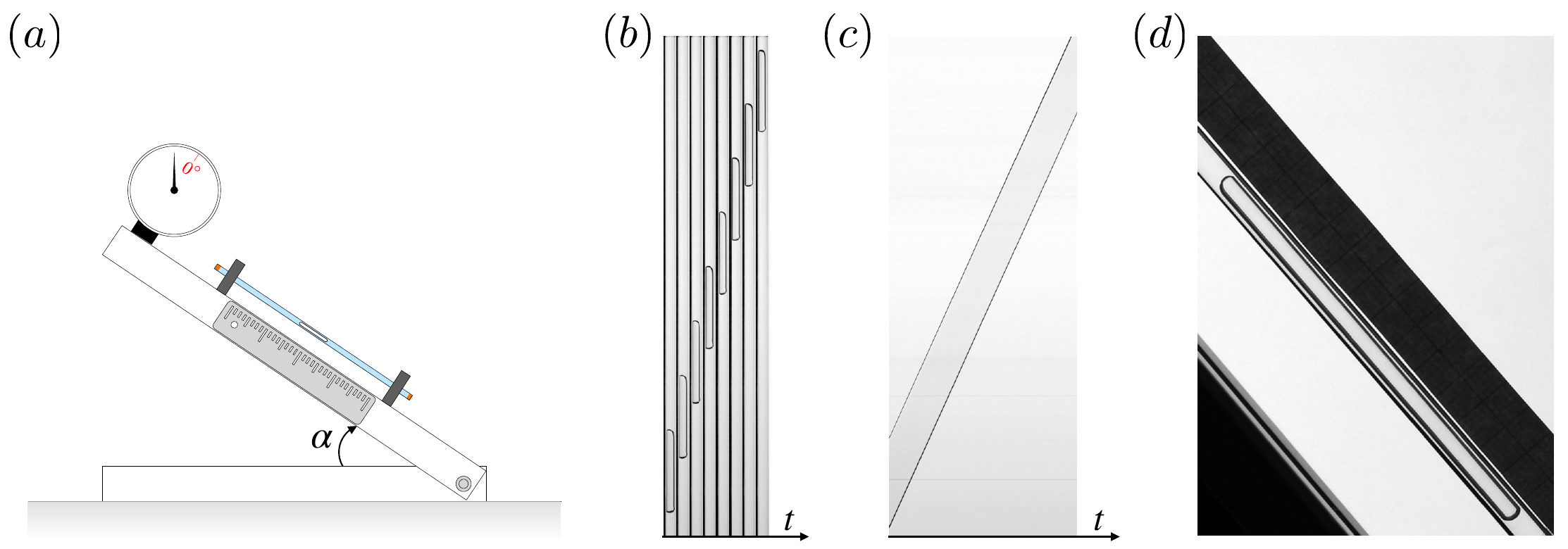}
\caption{(a) Sketch of the experimental setup. (b) Photographs of a long bubble inside a tube filled with silicone oil, at different and equally spaced time steps. Here, $\text{Bo}=1$ and the tube is tilted by $\alpha=35\degree$ with respect to the horizontal axis. (c) Intensity profile as a function of time along the tube axis. To produce this image, a column of pixel aligned with the central axis of the tube is extracted from each frame of the movie. The columns are then juxtaposed to each other. The locations of the upper and lower cap as a function of time are easily identified as the two roughy parallel black curves limiting a slightly darker domain that corresponds to the position of the bubble itself. The rising velocity is given by the slope of these black lines. As in (b), $\text{Bo}=1$ and $\alpha=35\degree$. (d) Photograph of a bubble in a tube tilted by $\alpha=50\degree$, with $\text{Bo}=0.7$. For these parameters, the system is close to critical conditions for the onset of motion.}
\label{set_up_inclined}
\end{figure}

The experimental setup is depicted on Figure \ref{set_up_inclined}(a). The tube is attached on an aluminium arm that can be tilted by an angle $\alpha \in \left[0\degree, 180\degree\right]$ with respect to the horizontal plane. A LED panel is positioned behind the setup for visualization purposes. Once the tilt angle is fixed, a camera records the rising motion of the bubble along the central axis of the tube. From the recorded footage, we can then retrieve the bubble velocity as previously described in Section \ref{sec:exp_centrifugated}, and as illustrated in Figure \ref{set_up_inclined}(b) and (c).  For the narrowest tubes where the bubble velocity is the smallest (if not zero), we use time-lapses instead of movies. 

Unlike the case of motor-driven rotating tubes, there is in principle no limitation on the observation time, allowing the detection of much slower bubble displacements. In practice, we consider the bubble velocity to be zero if the displacement of the bubble over a week is smaller than our resolution limit of 1mm. This implies that the smallest experimentally measurable capillary number is $\text{Ca}_\text{min}=7.6 \times10^{-9}$.

\subsection*{Experimental results and comparison with the theoretical prediction}

\begin{figure}
\centering
\includegraphics[width=\textwidth]{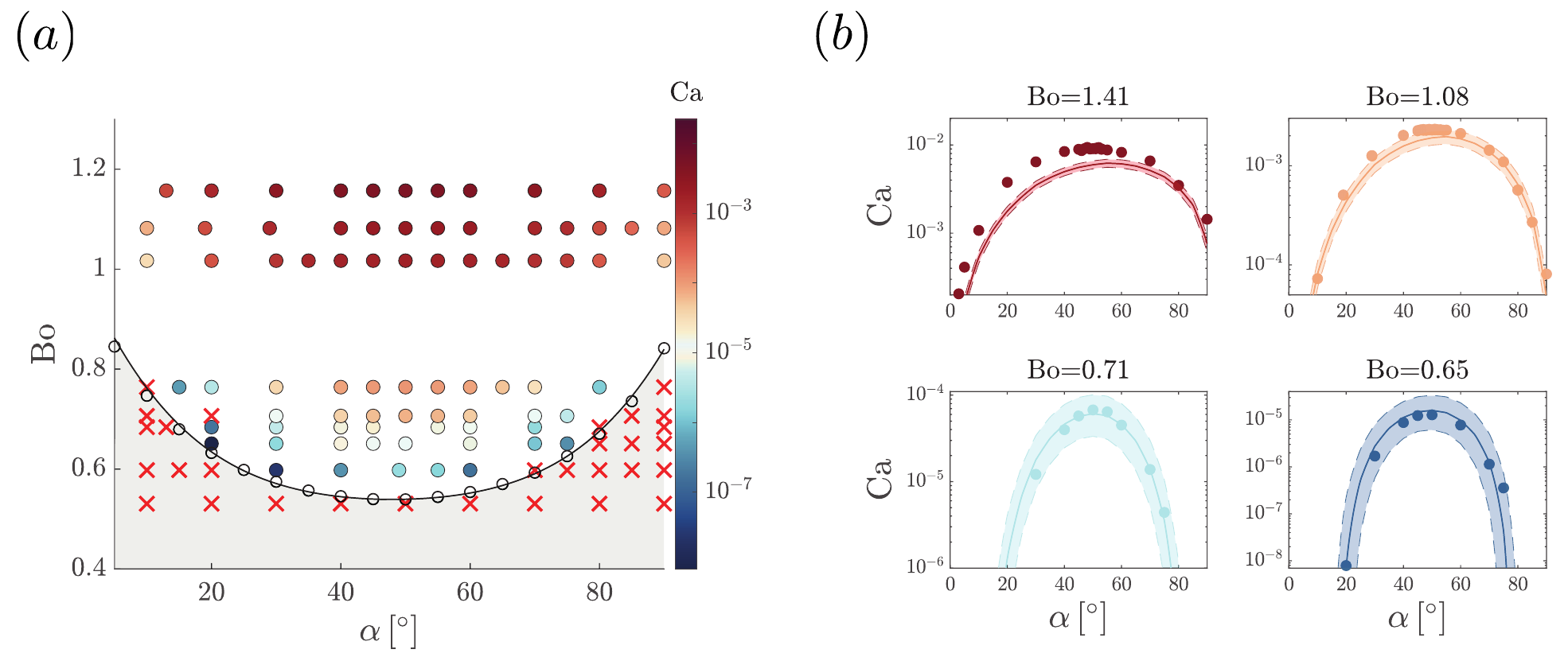}
\caption{(a) Diagram $(\text{Bo}, \alpha)$ where each dot corresponds to a measurement of $\text{Ca}>0$ for a given set of parameters $(\text{Bo}, \alpha)$. The red crosses indicate the couples $(\text{Bo}, \alpha)$ for which the bubble displacement fell below our detection limit. The black dots indicate the theoretical threshold for the onset of motion and the black solid line represents the polynomial approximation Eq.\eqref{eq:poly_approx_BoCr_vs_alpha}. Below this line, the gray area corresponds to the region of parameters where the steady rising of a bubble is not possible according to our theoretical analysis. (b) Velocity of the bubble as a function of the tilt angle $\alpha$ for various Bond numbers. The dots are the experimental points while for each Bond number, the solid line is described by Eq.\eqref{eq:prediction_velocity_inclined}, using the corresponding experimental value of Bo reported in the title of each panel. No fit parameter is used here: the values of $\beta(\alpha)$ and $\text{Bo}_c(\alpha)$ in Eq.\eqref{eq:prediction_velocity_inclined} are the ones displayed in Figure \ref{fig:phi_vs_alpha_Bo_vs_alpha_theorique}. The dotted lines also represent the prediction Eq.\eqref{eq:prediction_velocity_inclined}, but for $\text{Bo} \pm \Delta \text{Bo}$, where $\Delta$Bo accounts for the $\pm 0.05$ mm uncertainty on the tubes inner diameters. Here, the markers size represents the maximal measurement uncertainty on the bubble velocity.}
\label{exp_results_inclined}
\end{figure}

Our experimental findings are summarized and compared with our theoretical predictions in Figure \ref{exp_results_inclined}. Firstly, we observe that the bubble velocity strongly depends on the tilt angle $\alpha$ and reaches its maximum at approximately $\alpha \approx 50\degree$, a value independent of $\text{Bo}$ within the range of Bond numbers investigated here, as shown in Figure \ref{exp_results_inclined}(b). Furthermore, for $\text{Bo}_c(\alpha=50\degree)=0.5401<\text{Bo}<0.842=\text{Bo}_c(\alpha=90\degree)$, tilting the tube by the appropriate angle actually enables the motion of a bubble that would otherwise be stuck in a vertical configuration, as illustrated for instance by the cases $\text{Bo}=0.71$ and  $\text{Bo}=0.65$ reported in Figure \ref{exp_results_inclined}(b) . No motion at all is observed below the threshold $\text{Bo}_c(\alpha \approx 50\degree)$. Those observations align well with our theoretical analysis.

Finally, the bubble velocity as a function of the tilt angle $\alpha$ at low Bond numbers seems to be well described by Eq.\eqref{eq:prediction_velocity_inclined}, without any fitting parameter, see Figure \ref{exp_results_inclined}(b). We note that the agreement with the theoretical prediction appears to slightly deteriorate at larger Bond numbers. Indeed, several assumptions made in the theoretical analysis only hold in the vicinity of the threshold and are therefore expected to fail in the large $\text{Bo}$ regime. In large capillaries, the thin film thickness cannot be considered as uniform along the azimutal direction: the lubricating film is indeed much thicker in the direction $x>0$ \citep{Zukoski1966}. Similarly, requiring the static cap profile to expand in the entire fluid domain $r<R$ is likely to become inadequate as $\text{Bo}$ increases. All together however, the comparison tends to validate the relevance of a two-dimensional analysis to describe the thin lubricating film surrounding the bubble, even in a tilted configuration, in the low $\text{Bo}$ regime. 

It is worth mentioning that as a first attempt to describe the phenomenon, we opted for a fully two-dimensional description of the air-liquid interface. By matching the two-dimensional static cap with the thin film profile, we obtained the threshold $\text{Bo}_c^{2D}(\alpha)$ reported in Figure \ref{fig:phi_vs_alpha_Bo_vs_alpha_theorique}(b). 
This threshold exhibits the same non-monotonous trend as a function of the tilt angle, with a minimum reached for $\alpha_\text{opt} \lessapprox 50\degree$. However, it is downward-shifted with respect to the critical Bond number relying on a three-dimensional description of the static cap, which provides a much better agreement with experimental measurements, see Figure \ref{exp_results_inclined}(a). From this comparison, we conclude that while a simplified, two-dimensional description of the thin film region is acceptable, a proper characterization of the phenomenon requires to account for the three-dimensional shape of the static cap. 
\\

 We can now rationalize the theoretical and experimental results: the increase in transversal acceleration due to the tilt angle tends to increase the film thickness at the tongue of the static cap, enabling higher velocities within the tube for the same axial gravity. However, tilting the tube decreases the driving buoyancy force, which in turn reduces the bubble velocity. The interplay between these two effects leads to the observed non-monotonous dependency of the rising velocity on the tilt angle. In the limit case $\alpha \rightarrow 0\degree$ (horizontal tube), there is no motion within the tube since the driving force disappears.

\section{Conclusion}

In this study, we investigated theoretically and experimentally two different strategies aimed at enabling the motion of long air bubbles trapped in narrow, sealed capillaries partially filled with a viscous liquid. Both strategies, namely centrifugating the tube or tilting it with respect to its central axis, amount to modify the pressure distribution in the film surrounding the bubble by means of an external force field (centrifugal force or tilted gravity). This impacts both the shape of the static cap of the bubble, and the profile of the liquid-air interface in the thin film region. In particular, the resulting pressure gradients lead in both cases to the thickening of the lubricating film, thus enabling bubble ascent. The threshold for the onset of motion and the rising velocity above threshold as functions of the rotational speed and of tilt angle, respectively, are retrieved by the matching of the static cap and thin film profiles, that conditions the steady ascent of the bubble. Remarkably, the matching conditions in terms of film thickness and slope at the inflection point are in both cases the same as described in \cite{Bretherton1961} for the classical vertical setting (without rotation). However, both centrifugation and inclination alter the inner region solution and the static cap profile, making this matching possible for smaller Bond numbers. Thus, tunable parameters such as the rotational speed or the tilt angle, can effectively lower the threshold for the onset of motion, thus allowing the transport of bubbles even in very narrow capillaries.

The first part of this study was dedicated to the case of a vertical tube in rotation around its central symmetry axis. We extended Bretherton's analysis \citep{Bretherton1961} to account for the radial pressure gradient resulting from the tube centrifugation. By computing the shape of the tip of the bubble and solving the lubrication equation describing the thin film region, we could derive a matching condition yielding a theoretical prediction for the ascent velocity of the bubble, together with a new threshold for the onset of motion. Our theoretical findings highlight that centrifugating the tube acts as a downward shift on the critical bubble confinement. Our experimental campaign corroborated this analysis and confirmed the relevance of this strategy to release bubbles trapped in very narrow capillaries. 

In the second part, we explored how tilting the tube with respect to gravity could influence the transport of the bubble trapped inside. The three-dimensional static cap shape of the bubble was computed numerically, while the thin film region was assumed to be axisymmetric. By matching these profiles at the point of vanishing radial curvature, we could derive a prediction for the steady velocity of the bubble, that can only hold if the inner radius is larger than an angle-dependent critical value. This threshold varies non-monotonously with the tilt angle, with a minimum reached about $\alpha_\text{opt} \approx  48\degree$. Those predictions, although relying on a simplified description of the thin film region, align well with our experimental findings. 

Overall, these strategies seem well suited to many microfluidics applications where it is instrumental to get rid of trapped bubbles, without compromising the integrity of the capillary. The use of a tunable external force field provides a practical way to precisely monitor the motion of long bubbles. For further practical uses, we recall here the approximated expressions of the thresholds derived along this study: 
\begin{equation*}
\begin{aligned}
    \text{Bo}_c(\text{Ce}) &\approx 0.842-0.295 \, \text{Ce} + 0.020\,  \text{Ce}^2 \text{\, for centrifugated tubes, and:}\\
    \text{Bo}_{c}(\alpha)&\approx 0.54\left[1 + 0.5\left(\frac{\pi}{180}\right)^2(\alpha -48\degree)^2 + \left(\frac{\pi}{180}\right)^4(\alpha-48\degree)^4\right], \text{ for tilted tubes.}
\end{aligned}
\end{equation*}

At a more fundamental level, these strategies provide an interesting framework to examine the infinitely slow dynamics of pinch-off, a phenomenon explored theoretically  by \cite{Lamstaes2017}  and experimentally investigated by \cite{Dhaouadi2019} in capillary tubes with inner radii $R<R_c$. For instance, starting from a moving bubble within a rotating capillary and subsequently halting the rotation offers a practical means to establish a precisely defined initial condition, from which the pinching process starts.

\subsubsection*{\textbf{\textup{Funding}}}
\indent \, We acknowledge the Swiss National Science Foundation under grant 200341.
\subsubsection*{\textbf{\textup{Declaration of Interests}}}
\indent \, The authors report no conflict of interest.

\begin{appendix}
    \section{Derivation of the thin film velocity profile for the centrifugal and tilted case}
    
\subsection{Thin film velocity profile in a centrifugated tube}
\label{app:derivation_velocity_centrigated}
Here, we derive the velocity profile in the thin film region when the (vertical) tube is rotated around its central axis at angular frequency $\omega$. 
In the rotating reference frame, translating with the bubble at steady velocity $U_b \mathbf{e}_z$, the stationary Navier-Stokes equations write: 
\begin{equation}
\boldsymbol{\nabla}.\boldsymbol{u}=0, \quad
\rho \left[\left(\boldsymbol{u}.\boldsymbol{\nabla}\right)\boldsymbol{u}+2 \boldsymbol{\Omega }\times \boldsymbol{u}\right]=-\boldsymbol{\nabla}p- \rho \boldsymbol{\Omega}\times( \boldsymbol{\Omega} \times \boldsymbol{r})+\mu \Delta \boldsymbol{u} + \rho \boldsymbol{g},
\end{equation}

\noindent where $\mathbf{\Omega}=\omega \mathbf{e}_z$ is the rotation vector. 

We introduce the dimensionless variables $\bar{\boldsymbol{u}}$, $\bar{z}$, and $\bar{r}$, and $\bar{p}$ , such that $\boldsymbol{u}=U_b\bar{\boldsymbol{u}}$, $z=R\bar{z}$,  $r=R\bar{r}$ and $p=\pi \bar{p}$, where $\pi= \rho \omega R U_b$. The dimensionless Navier-Stokes equations read:
\begin{equation}
\boldsymbol{\nabla}.\bar{\boldsymbol{u}}=0,\quad
\text{Ro} \left(\bar{\boldsymbol{u}}.\boldsymbol{\nabla}\right)\bar{\boldsymbol{u}} + 2 \boldsymbol{e}_z \times \bar{\boldsymbol{u}}=-\boldsymbol{\nabla}\bar{p}- \frac{1}{\text{Ro}} \boldsymbol{e}_z\times( \boldsymbol{e}_z \times \bar{\boldsymbol{r}})+\text{E} \Delta \bar{\boldsymbol{u}} - \frac{g}{\omega U_b}\boldsymbol{e}_z,
\end{equation}
\noindent where $\text{Ro}= \frac{U_b}{\omega R}$ is the Rossby number and $\text{E}=\frac{\nu }{\omega R^2}$ is the Ekman number. In the experiments presented in this study, $\text{Ro}\ll 1$ and the non-linear terms of the Navier-Stokes equation can therefore be neglected. Under these assumptions, and enforcing axisymmetry,  the stationary Navier-Stokes equations in cylindrical coordinates write:  

\begin{align}
0&=\frac{1}{r} \frac{\partial}{\partial r}\bigg( r u_r\bigg) + \frac{\partial u_z}{\partial z},\\
-2 \rho \omega u_\theta&=-\pd{p}{r}+\rho \omega^2 r+\mu\left[\pd{}{r}\left(\frac{1}{r}\pd{}{r}(ru_r)\right)+\frac{\partial^2 u_r}{\partial z^2}\right],\\
2\rho \omega u_r&=\mu\left[\pd{}{r}\left(\frac{1}{r}\pd{}{r}(ru_\theta)\right)+\frac{\partial^2u_\theta}{\partial z^2} \right],\\
0&=-\pd{p}{z}+\mu\left[\frac{1}{r}\pd{}{r}\left(r\pd{u_z}{r}\right)+\frac{\partial^2u_z}{\partial z^2}\right]-\rho g.
\end{align}

Furthermore, in cylindrical coordinates, the stress tensor writes: 
\begin{equation*}
\boldsymbol{\sigma}=\begin{pmatrix}
-p+2\mu \pd{u_r}{r} & \mu r\pd{}{r}\left(\frac{u_\theta}{r}\right)&\mu\left(\pd{u_r}{z}+\pd{u_z}{r}\right)\\
 \mu r\pd{}{r}\left(\frac{u_\theta}{r}\right) & -p+\frac{2\mu}{r}u_r & \mu \pd{u_\theta}{z} \\
 \mu\left(\pd{u_r}{z}+\pd{u_z}{r}\right) &  \mu \pd{u_\theta}{z} & -p+2\mu \pd{u_z}{z}
\end{pmatrix}, 
\end{equation*}

\noindent and the vector normal to the interface is $\boldsymbol{n}=\frac{1}{\sqrt{1+r'_1(z)^2}}\left(-1, 0, r'_1(z)\right)^T$.

\noindent Therefore, the dynamic and kinematic boundary conditions at the fluid-air interface are
\begin{align}
\gamma \kappa&=\left(p-p_\text{{air}}\right)\left(1+r'_1(z)^2\right)-2\mu \left[\pd{u_r}{r}+r'_1(z)^2\pd{uz}{z}\right]+2\mu r'_1(z)\left[\pd{u_r}{z}+\pd{u_z}{r}\right],\\
0&=2r'_1(z)\left[\pd{u_r}{r}-\pd{u_z}{z}\right]+\left(1-r'_1(z)^2\right)\left[\pd{u_r}{z}+\pd{u_z}{r}\right],
\end{align}
\noindent where $\kappa=-\frac{1}{r_1(z)\sqrt{1+r'_1(z)^2}}+\frac{r''_1(z)}{\left(1+r'_1(z)^2\right)^{3/2}}$ is the curvature. 

Finally, the no-slip boundary condition at the solid wall implies $u_z(r=R)=-U_b$.

\subsection*{Change of coordinates}

\begin{figure}
    \centering
    \includegraphics[width=0.5\textwidth]{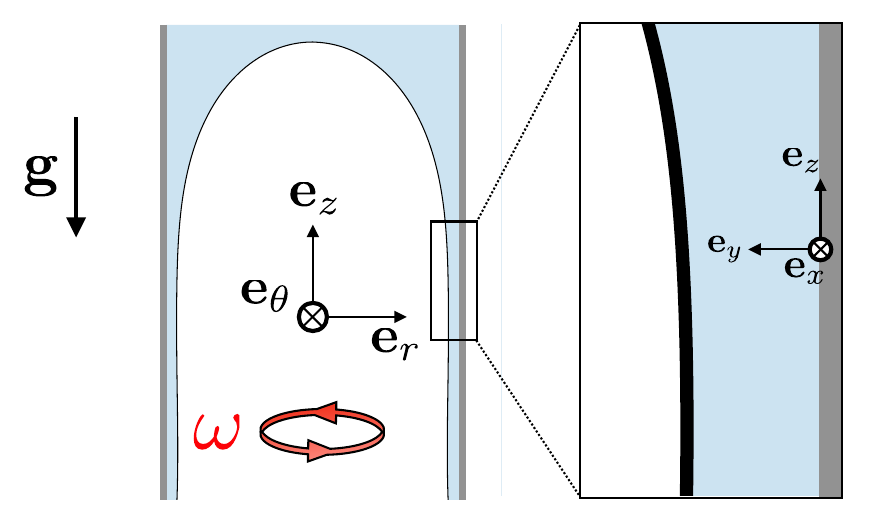}
    \caption{Sketch of the upper part of a Taylor bubble in a sealed vertical tube, rotating around its symmetry axis at angular velocity $\omega$. The flow around the bubble is first described in cylindrical coordinates $(r, \theta, z)$, where $\mathbf{e}_z$ is aligned with the tube axis. We focus on the thin film region, that can be considered as planar instead of annular, and describe then the lubricating film in Cartesian coordinates $(x, y=R-r, z)$.  }
    \label{fig:appendix_centrifugated}
\end{figure}

In the thin film region, the film thickness is very small compared to the radius, so that the flow can be treated as if the region were planar, instead of annular. Accordingly, we describe the flow in the Cartesian  coordinate system $(x, y=R-r, z)$, see Figure \ref{fig:appendix_centrifugated}. Furthermore, we introduce the modified pressure field: $P=p+\rho \omega^2 Ry + \rho g z$. In this system of coordinates, the Navier-Stokes equation become: 

\begin{align}
0&=\frac{\partial u_y}{\partial y}+\frac{\partial u_z}{\partial z} -\frac{u_y}{R-y},\\
-2\rho \omega u_x &=\frac{\partial P}{\partial y}-\rho \omega^2y+\mu \left[\frac{1}{R-y}\frac{\partial u_y}{\partial y}+\frac{u_y}{(R-y)^2}-\frac{\partial ^2u_y}{\partial y^2}   -\frac{\partial^2u_y}{\partial z^2}\right],\\
-2\rho \omega u_y&=\mu\left[- \frac{1}{R-y}\frac{\partial u_x}{\partial y}-\frac{u_x}{(R-y)^2} + \frac{\partial^2u_x}{\partial y^2}+\frac{\partial^2u_x}{\partial z^2}\right],\\
0&=-\frac{\partial P}{\partial z} + \mu\left[ -\frac{1}{R-y}\frac{\partial u_z}{\partial y}+\frac{\partial ^2u_z}{\partial y^2}+\frac{\partial^2 u_z}{\partial z^2}\right] .
\end{align}

Likewise, the dynamic and kinematic boundary conditions become: 
\begin{align}
\gamma \kappa&=\left(P-P_\text{air}\right)\left(1+y_1'(z)^2\right)-2\mu \left[\pd{u_y}{y}+y_1'(z)^2\pd{uz}{z}\right]+2\mu y_1'(z)\left[\pd{u_y}{z}+\pd{u_z}{y}\right],\\
0&=2y_1'(z)\left[\pd{u_y}{y}-\pd{u_z}{z}\right]+\left(1-y_1'(z)^2\right)\left[\pd{u_y}{z}+\pd{u_z}{y}\right],
\end{align}

\noindent where $\kappa=-\frac{1}{(R-y_1(z))\sqrt{1+y_1'(z)^2}}-\frac{y_1''(z)}{\left(1+y_1'(z)^2\right)^{3/2}}$.

\subsection*{Lubrication approximation}

We adimensionalize as follows: $u_z=U_b \overline{u_z}$, $u_y=U_y \overline{u_y}$, $u_x=U_x \overline{u_x}$, $P=P_0 \overline{P}$, $y=b\overline{y}$, $y_1=b\overline{y_1}$, $z=R\overline{z}$, where $\epsilon =\frac{b}{R} \ll 1$. According to the least degeneracy principle applied to the mass conservation equation, $U_y=\epsilon U_b$ and the mass conservation equation becomes: $\pd{\overline{u_y}}{\overline{y}}+\pd{\overline{u_z}}{\overline{z}}=0$.

Furthermore, upon introduction of the dimensionless fields and variables, the momentum conservation equations are written as: 

\begin{align}
-2\rho \omega U_x \overline{u_x} &=\frac{P_0}{b}\frac{\partial \overline{P}}{\partial \overline{y}}-\rho \omega^2b \bar{y}\, + \nonumber\\
&\quad \quad \quad \quad \epsilon \frac{\mu U_b}{b^2} \left[\frac{\epsilon}{1-\epsilon\overline{y}}\frac{\partial \overline{u_y}}{\partial \overline{y}}+\epsilon^2\frac{\overline{u_y}}{(1-\epsilon \overline{y})^2}-\frac{\partial ^2\overline{u_y}}{\partial \overline{y}^2}   -\epsilon^2\frac{\partial^2\overline{u_y}}{\partial \overline{z}^2}\right],\\
-2\rho \omega \epsilon U_b \overline{u_y}&=\frac{\mu U_x}{b^2}  \left[- \frac{\epsilon}{1-\epsilon\overline{y}}\frac{\partial \overline{u_x}}{\partial \overline{y}}-\epsilon^2\frac{\overline{u_x}}{(1-\epsilon \overline{y})^2}+ \frac{\partial^2\overline{u_x}}{\partial \overline{y}^2}+\epsilon^2\frac{\partial^2\overline{u_x}}{\partial \overline{z}^2}\right],\\
0&=-\frac{P_0}{R} \frac{\partial \overline{P}}{\partial \overline{z}} + \frac{\mu U_b}{b^2}\left[ - \frac{\epsilon}{1-\epsilon \overline{y}}\frac{\partial \overline{u_z}}{\partial \overline{y}}+\frac{\partial ^2\overline{u_z}}{\partial \overline{y}^2}+\epsilon^2\frac{\partial^2 \overline{u_z}}{\partial \overline{z}^2}\right].
\end{align}

The least degeneracy principle applied to the momentum conservation equations along the $x$ and $z$- axis implies that $U_x=2 \epsilon \rho \omega U_b b^2/\mu$ and $P_0=\frac{\mu U_b}{b\epsilon}$.

Thus, the momentum conservation equations along the $y$-axis becomes : 

\begin{equation}
-4\epsilon \text{Re} \overline{u_x} = \frac{\text{Ca}}{\epsilon^4\text{Ce}}\frac{\partial \overline{P}}{\partial \overline{y}}-\overline{y}+
\frac{\text{Ca}}{\epsilon^2\text{Ce}}  \left[\frac{\epsilon}{1-\epsilon\overline{y}}\frac{\partial \overline{u_y}}{\partial \overline{y}}+\epsilon^2\frac{\overline{u_y}}{(1-\epsilon\overline{y})^2}-\frac{\partial ^2\overline{u_y}}{\partial \overline{y}^2}   -\epsilon^2\frac{\partial^2\overline{u_y}}{\partial \overline{z}^2}\right]
\end{equation}
\noindent with $\text{Re}=\rho U_b b/\mu$, $\text{Ca}=\mu U_b/\gamma$ and $\text{Ce}=\rho \omega^2 R^3/\gamma$. From the volume conservation constraint $\rho g b^3/3\mu U_b=R/2$, we deduce that $\frac{\text{Ca}}{\text{Ce}}\sim  \epsilon^3 \frac{2g}{3\omega^2R} = O(\epsilon^3)$. Thus, the system of equations reduces to: 
\begin{equation}
0=\pd{\overline{u_y}}{\overline{y}}+\pd{\overline{u_z}}{\overline{z}}, \quad \quad
0 = \frac{\partial \overline{P}}{\partial \overline{y}}, \quad \quad
0=- \frac{\partial \overline{P}}{\partial \overline{z}} +\frac{\partial ^2\overline{u_z}}{\partial \overline{y}^2}.
\end{equation}

The dynamic and kinematic boundary conditions are: 
\begin{align}
\gamma \kappa&=\frac{\mu U_b}{b\epsilon} \left(\overline{P}-\overline{P}_\text{air}\right)\left(1+\epsilon^2 \overline{y_1}'(\overline{z})^2\right)-2\epsilon \frac{\mu U_b}{b} \left[\pd{\overline{u_y}}{\overline{y}}+\epsilon^2 \overline{y_1}'(\overline{z})^2\pd{\overline{u_z}}{\overline{z}}\right]\nonumber\\
&\quad \quad \quad \quad \quad \quad \quad + 2 \epsilon \frac{\mu U_b}{b}\overline{y_1}'(\overline{z})\left[\epsilon^2 \pd{\overline{u_y}}{\overline{z}}+\pd{\overline{u_z}}{\overline{y}}\right],\\
0&=2\epsilon^2 \overline{y_1}'(\overline{z})\left[\pd{\overline{u_y}}{\overline{y}}-\pd{\overline{u_z}}{\overline{z}}\right]+\left(1-\epsilon^2 \overline{y_1}'(\overline{z})^2\right)\left[\epsilon^2 \pd{\overline{u_y}}{\overline{z}}+\pd{\overline{u_z}}{\overline{y}}\right].
\end{align}

Thus, including the no-slip boundary condition at the solid wall, the boundary conditions at leading order are: 
\begin{equation}
\overline{P}-\overline{P}_\text{air}=\gamma \kappa \frac{\epsilon b}{\mu U_b},\quad \quad
\frac{\partial \overline{u_z}}{\partial \overline{y}}=0,\quad \quad \overline{u_z}(\overline{y}=0)=-1. 
\end{equation}

Finally, going back to the dimensional form, and reintroducing the original pressure field $p=P-\rho \omega^2 Ry - \rho g z$, the full problem writes: 
\begin{equation}
0=\pd{u_y}{y}+\pd{u_z}{z}, \quad \quad \pd{p}{y}=-\rho \omega^2 R, \quad \quad
\pd{p}{z}=\mu \frac{\partial ^2 u_z}{\partial y^2} - \rho g,\\
\end{equation}
\noindent and is complemented by the following boundary conditions: 
\begin{equation}
p(y=y_1, z)-p_\text{air}=\gamma \kappa, \quad \quad \pd{u_z}{y}\vert_{y=y_1}=0, \quad \quad
u_z(y=0, z)=-U_b.
\end{equation}

\noindent The integration of the pressure field is straightforward and leads to:
\begin{equation}
p(y, z)=p_\text{air} + \gamma \kappa + \rho \omega^2R\left(y_1(z)-y\right).
\end{equation}

\noindent Finally, by injecting this pressure field in the axial component of the momentum equation, we can derive the following equation for the velocity in the thin film: 
\begin{equation}
\mu \frac{\partial^2 u_z}{\partial y^2}=\gamma \kappa' +\rho \omega^2 Ry_1'+\rho g,
\end{equation}
\noindent that results into: 
\begin{equation}
u_z(y, z)=-U_b+ \frac{\gamma}{2 \mu} \left(\kappa' +\frac{\rho \omega^2 R}{\gamma} y_1' +\frac{\rho g}{\gamma}\right)(y^2-2y_1y). 
\end{equation}

\subsection{Thin film velocity profile in a tilted tube}
\label{app:derivation_velocity_tilted}

\begin{figure}
    \centering
    \includegraphics[width=0.7\textwidth]{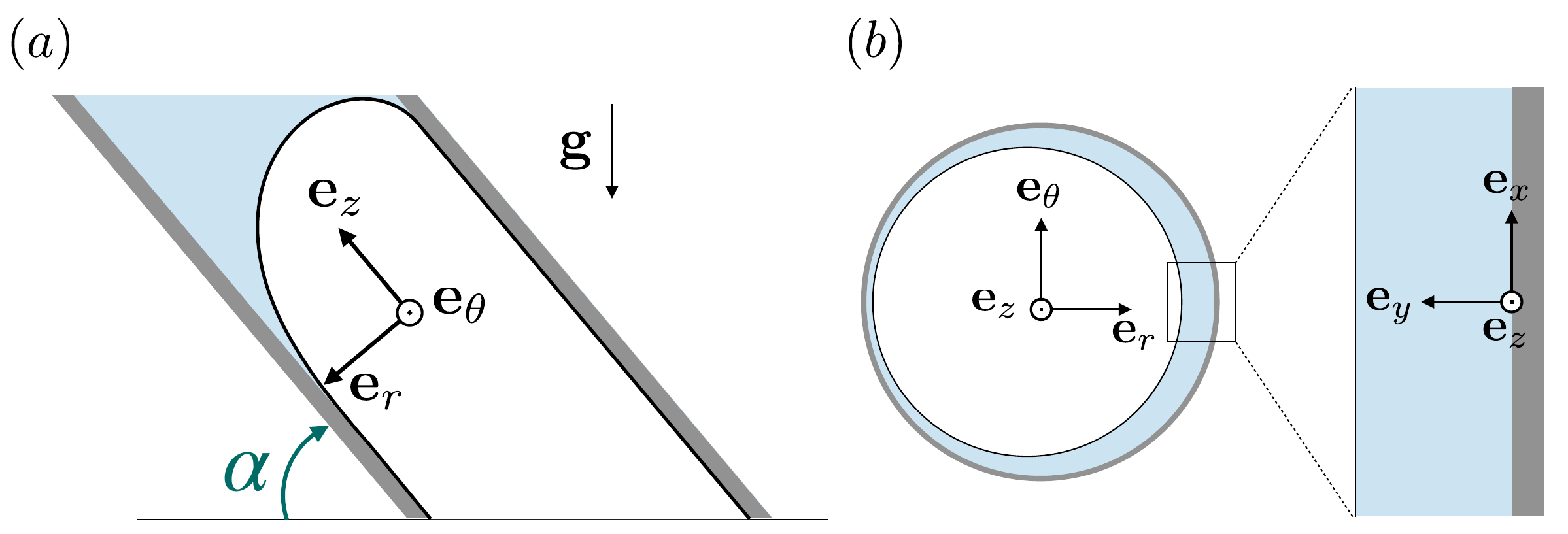}
    \caption{(a) Sketch of the upper part of a Taylor bubble in a sealed tube tilted by an angle $\alpha$ with respect to the horizontal plane. The flow in the thin lubricating film is described in cylindrical coordinates $(r, \theta, z)$,  where $\mathbf{e}_z$ is the direction aligned with the tube axis, such that $\mathbf{e}_z\cdot \mathbf{g}=-g\sin(\alpha)$. The origin of $\theta$ is chosen such that $\mathbf{e}_r(\theta=0)\cdot \mathbf{g}=g\cos(\alpha)$. (b) Sketch of the cross-section of the channel and of the bubble. We focus on the region in the vicinity of the plane $\theta=0$, described in Cartesian coordinates $(x, y=R-r, z)$, where $\mathbf{e}_x\cdot \mathbf{g}=0$, $\mathbf{e}_y \cdot \mathbf{g}=-g\cos(\alpha)$, and $\mathbf{e}_z \cdot \mathbf{g}=-g\sin(\alpha)$.}
    \label{fig:fig_appendix}
\end{figure}

We now aim at deriving the velocity profile in the thin film region when the tube is tilted by an angle $\alpha$.
Since the Reynolds number $\text{Re}= {\rho U_b b}/{\mu}$ characterizing the flow in the thin film region is very small, we can safely neglect the effect of inertia. In the stationary cylindrical system of coordinates $(r, \theta, z)$, translating with the bubble at steady velocity $U_b \mathbf{e}_z$ where $z$ is aligned with the central axis of the (tilted) tube, see Figure \ref{fig:fig_appendix}(a), the Navier-Stokes equation read:

\begin{align}
0&=\frac{1}{r} \frac{\partial}{\partial r}\bigg( r u_r\bigg) + \frac{1}{r} \frac{\partial u_\theta}{\partial \theta}+ \frac{\partial u_z}{\partial z},\\
0&=-\pd{p}{r}+\mu\left[\pd{}{r}\left(\frac{1}{r}\pd{}{r}(ru_r)\right)+\frac{1}{r^2} \frac{\partial^2 u_r}{\partial \theta^2} - \frac{2}{r^2} \frac{\partial u_\theta}{\partial \theta} +\frac{\partial^2 u_r}{\partial z^2}\right] + \rho g \cos(\theta) \cos(\alpha),\\
0&=-\pd{p}{\theta} + \mu\left[\pd{}{r}\left(\frac{1}{r}\pd{}{r}(ru_\theta)\right)+  \frac{1}{r^2} \frac{\partial^2 u_\theta}{\partial \theta^2} +\frac{2}{r^2} \frac{\partial u_r}{\partial \theta}  + \frac{\partial^2u_\theta}{\partial z^2} \right]- \rho g \sin(\theta)\cos(\alpha),\\
0&=-\pd{p}{z}+\mu\left[\frac{1}{r}\pd{}{r}\left(r\pd{u_z}{r}\right)+\frac{1}{r^2}\frac{\partial^2 u_z}{\partial \theta^2} + \frac{\partial^2u_z}{\partial z^2}\right]-\rho g\sin(\alpha).
\end{align}

We know from our analysis of the three-dimensional cap profile, that the matching with the thin film region profile should be imposed at the tip of the tongue exhibited by the static cap, i.e. at the point of coordinates $(r=R, \theta=0, h(R, 0))$. We will thus restrict the study of the thin film solution to the plane $(\theta=0)$, see Figure \ref{fig:fig_appendix}(b). By assuming an vanishing azimuthal curvature $\sim 1/R$, the region of size $\sim R\text{d}\theta$ in the close vicinity of $\theta=0$ can be considered infinite. Yet, in the vicinity of $\theta=0$, the derivative with respect to $\theta$ should vanish by symmetry. Therefore, in this region, the Navier-Stokes equations reduce to: 
\begin{align}
0&=\frac{1}{r} \frac{\partial}{\partial r}\bigg( r u_r\bigg) +\frac{\partial u_z}{\partial z},\\
0&=-\pd{p}{r}+\mu\left[\pd{}{r}\left(\frac{1}{r}\pd{}{r}(ru_r)\right) + \frac{\partial^2 u_r}{\partial z^2}\right] + \rho g \cos(\alpha),\\
0&= \mu\left[\pd{}{r}\left(\frac{1}{r}\pd{}{r}(ru_\theta)\right)+ \frac{\partial^2u_\theta}{\partial z^2} \right],\\
0&=-\pd{p}{z}+\mu\left[\frac{1}{r}\pd{}{r}\left(r\pd{u_z}{r}\right)+ \frac{\partial^2u_z}{\partial z^2}\right]-\rho g\sin(\alpha).
\end{align}

These equations are complemented by the following dynamic and kinematic boundary conditions: 
\begin{align}
\gamma \kappa&=\left(p-p_\text{{air}}\right)\left(1+r'_1(z)^2\right)-2\mu \left[\pd{u_r}{r}+r'_1(z)^2\pd{uz}{z}\right]+2\mu r'_1(z)\left[\pd{u_r}{z}+\pd{u_z}{r}\right],\\
0&=2r'_1(z)\left[\pd{u_r}{r}-\pd{u_z}{z}\right]+\left(1-r'_1(z)^2\right)\left[\pd{u_r}{z}+\pd{u_z}{r}\right],
\end{align}
\noindent where $\kappa=-\frac{1}{r_1(z)\sqrt{1+r'_1(z)^2}}+\frac{r''_1(z)}{\left(1+r'_1(z)^2\right)^{3/2}}$ is the curvature, and by the no-slip boundary condition at the solid wall: $u_z(r=R)=-U_b$.

\subsection*{Change of coordinates}

As before, we neglect curvature in the azimuthal direction, and describe the thin film region within the Cartesian  coordinate system $(x, y=R-r, z)$, see Figure \ref{fig:fig_appendix}(b). Furthermore, we introduce the modified pressure field: $P=p + \rho g \cos(\alpha)y + \rho g \sin(\alpha)z$. In this system of coordinates, the mass conservation and momentum conservation equations along the $y$ and $z$ directions become: 
\begin{align*}
0&=\frac{\partial u_y}{\partial y}+\frac{\partial u_z}{\partial z} -\frac{u_y}{R-y},\\
0 &=\frac{\partial P}{\partial y}+\mu \left[\frac{1}{R-y}\frac{\partial u_y}{\partial y}+\frac{u_y}{(R-y)^2}-\frac{\partial ^2u_y}{\partial y^2}   -\frac{\partial^2u_y}{\partial z^2}\right],\\
0&=-\frac{\partial P}{\partial z} + \mu\left[ -\frac{1}{R-y}\frac{\partial u_z}{\partial y}+\frac{\partial ^2u_z}{\partial y^2}+\frac{\partial^2 u_z}{\partial z^2}\right].
\end{align*}

Likewise, the dynamic and kinematic boundary conditions become: 
\begin{align}
\gamma \kappa&=\left(P-P_\text{air}\right)\left(1+y_1'(z)^2\right)-2\mu \left[\pd{u_y}{y}+y_1'(z)^2\pd{uz}{z}\right]+2\mu y_1'(z)\left[\pd{u_y}{z}+\pd{u_z}{y}\right],\\
0&=2y_1'(z)\left[\pd{u_y}{y}-\pd{u_z}{z}\right]+\left(1-y_1'(z)^2\right)\left[\pd{u_y}{z}+\pd{u_z}{y}\right],
\end{align}

\noindent where $\kappa=-\frac{1}{(R-y_1(z))\sqrt{1+y_1'(z)^2}}-\frac{y_1''(z)}{\left(1+y_1'(z)^2\right)^{3/2}}$. 

\subsection*{Lubrication approximation}
We adimensionalize as follows: $u_z=U_b \overline{u_z}$, $u_y=U_y \overline{u_y}$, $P=P_0 \overline{P}$, $y=b\overline{y}$, $y_1=b\overline{y_1}$, $z=R\overline{z}$, where $\epsilon =\frac{b}{R} \ll 1$. According to the least degeneracy principle applied to the mass conservation equation, $U_y=\epsilon U_b$ and the mass conservation equation becomes: $0=\pd{\overline{u_y}}{\overline{y}}+\pd{\overline{u_z}}{\overline{z}}$.

Furthermore, upon introduction of the dimensionless fields and variables, the momentum conservation equations along the $y$ and $z$ directions are written as: 

\begin{align}
0&=\frac{P_0}{b}\pd{\overline{P}}{\overline{y}} + \frac{\epsilon \mu U_b}{b^2} \left[\frac{\epsilon}{1-\epsilon \overline{y}} \pd{\overline{u_y}}{\overline{y}} + \frac{\epsilon^2}{(1-\epsilon \overline{y})^2}\overline{u_y} -\frac{\partial^2 \overline{u_y}}{\partial \overline{y}^2} - \epsilon^2 \frac{\partial^2\overline{u_y}}{\partial \overline{z}^2} \right],\\
0&=-\frac{P_0}{R} \pd{\overline{P}}{\overline{z}} + \frac{ \mu U_b}{b^2} \left[ -\frac{\epsilon}{1-\epsilon \overline{y}}\pd{\overline{u_z}}{\overline{y}}+\frac{\partial^2 \overline{u_z}}{\partial \overline{y}^2}  +\epsilon^2 \frac{\partial^2 \overline{u_z}}{\partial \overline{z}^2}  \right].
\end{align}

The least degeneracy principle applied to the momentum conservation equation along the $z$-axis implies that: $P_0=\frac{\mu U_b}{b\epsilon}$. At leading order, the problem reduces then to: 
\begin{equation}
0=\pd{\overline{u_y}}{\overline{y}}+\pd{\overline{u_z}}{\overline{z}}, \quad \quad
0=\pd{\overline{P}}{\overline{y}}, \quad \quad
0=- \pd{\overline{P}}{\overline{z}} +  \frac{\partial^2 \overline{u_z}}{\partial \overline{y}^2}. 
\end{equation}

The dynamic and kinematic boundary conditions are: 
\begin{align}
\gamma \kappa&=\frac{\mu U_b}{b\epsilon} \left(\overline{P}-\overline{P}_\text{air}\right)\left(1+\epsilon^2 \overline{y_1}'(\overline{z})^2\right)-2\epsilon \frac{\mu U_b}{b} \left[\pd{\overline{u_y}}{\overline{y}}+\epsilon^2 \overline{y_1}'(\overline{z})^2\pd{\overline{u_z}}{\overline{z}}\right]\nonumber\\
&\quad \quad \quad \quad + 2 \epsilon \frac{\mu U_b}{b}\overline{y_1}'(\overline{z})\left[\epsilon^2 \pd{\overline{u_y}}{\overline{z}}+\pd{\overline{u_z}}{\overline{y}}\right],\\
0&=2\epsilon^2 \overline{y_1}'(\overline{z})\left[\pd{\overline{u_y}}{\overline{y}}-\pd{\overline{u_z}}{\overline{z}}\right]+\left(1-\epsilon^2 \overline{y_1}'(\overline{z})^2\right)\left[\epsilon^2 \pd{\overline{u_y}}{\overline{z}}+\pd{\overline{u_z}}{\overline{y}}\right].
\end{align}

Therefore at leading order, and including the no-slip boundary condition at the solid wall, the boundary conditions write: 
\begin{equation}
\overline{P}-\overline{P}_\text{air}=\gamma \kappa \frac{\epsilon b}{\mu U_b},\quad \quad 
\frac{\partial \overline{u_z}}{\partial \overline{y}}=0, \quad \quad \overline{u_z}(\overline{y}=0)=-1.
\end{equation}

Finally, going back to the dimensional form, and reintroducing the original pressure field $p=P-\rho g \cos(\alpha)y - \rho g \sin(\alpha)z$, the full problem reduces to: 
\begin{equation}
0=\pd{u_y}{y}+\pd{u_z}{z}, \quad \quad \pd{p}{y}=-\rho g\cos(\alpha), \quad \quad 
\pd{p}{z}=\mu \frac{\partial ^2 u_z}{\partial y^2} - \rho g \sin(\alpha), 
\end{equation}
\noindent and is complemented by the following boundary conditions: 
\begin{equation}
p(y=y_1, z)-p_\text{air}=\gamma \kappa, \quad \quad \pd{u_z}{y}\vert_{y=y_1}=0, \quad \quad 
u_z(y=0, z)=-U_b.
\end{equation}

\noindent The pressure field integrates straightforwardly into:
\begin{equation}
p(y, z)=p_\text{air} + \gamma \kappa + \rho g \cos(\alpha) \left(y_1(z)-y\right).
\end{equation}

\noindent Finally, by injecting this pressure field in the axial component of the momentum equation, we can derive the following equation for the velocity in the thin film: 
\begin{equation}
\mu \frac{\partial^2 u_z}{\partial y^2}=\gamma \kappa' +\rho g \cos(\alpha) y_1'+\rho g\sin(\alpha),
\end{equation}
\noindent which leads to the velocity profile: 
\begin{equation}
u_z(y, z)=-U_b+ \frac{\gamma}{2 \mu} \left(\kappa' +\frac{\rho g \cos(\alpha)}{\gamma} y_1' +\frac{\rho g\sin(\alpha)}{\gamma}\right)(y^2-2y_1y). 
\end{equation}

\section{Derivation of the equilibrium equation for the static cap}
\label{app:young_laplace_equation}

Following previous works (see e.g. \cite{Rascon2017,lubbers2014, Manning2011}), we derive from energy principles the three-dimensional equilibrium equation for the equilibrium of the static cap.
We introduce the coordinate system $(x,y,z)$, with the $z$ axis aligned with the central tube axis, and the $x$ axis aligned with the gravity component normal to the $z$ axis (so that $\mathbf{e}_x.\mathbf{g}=g\cos(\alpha)$, see Figure \ref{fig:config_validation_inclined}(a)). The location of the air-liquid interface is denoted by $h(x,y)$. 

The Gibbs free energy associated with the cap interface can be written as:
\begin{equation}
    E(h)=\gamma \mathcal{A} +\mathcal{G}, 
\end{equation}
\noindent where the first term on the  right-hand-side is the surface energy. As in \cite{Rascon2017}, the surface $\mathcal{A}$ is computed as 
\begin{equation}
    \mathcal{A} =\int_\Omega \sqrt{1+(\nabla h)^2}\,dx dy ,
\end{equation}
\noindent with $\Omega$ the cross-section of the capillary. 
The term $\mathcal{G}$ represents in turn the gravitational potential energy, that in general form reads \citep{pitts1973}: 
\begin{equation}
    \mathcal{G}=- \int_V \rho \boldsymbol{g} \cdot \boldsymbol{r} \,dx dy dz, \quad
    \boldsymbol{r}=(x,y,z), \quad \boldsymbol{g}=g (\cos \alpha, 0, -\sin \alpha), 
\end{equation}
\noindent where the volume $V$ is given by: 
\begin{equation}
V=\int_\Omega dx dy h(x,y).
\end{equation}
Upon introduction of the Lagrange multiplier $\lambda$ to ensure volume conservation, the functional to be minimized to obtain equilibrium reads:
\begin{equation}
    F(h)=\gamma \int_\Omega \sqrt{1+(\nabla h)^2}\,dx dy + \rho g \int_V \left( - x \cos \alpha + z \sin \alpha \right)\, dx dy dz - \lambda \int_V dx dy dz.
\end{equation}
Upon integration along the $z$ direction between 0 and $h$:
\begin{equation}
     F(h)=\gamma \int_\Omega \sqrt{1+(\nabla h)^2}\,dx dy + \rho g \int_\Omega \left( - x \cos \alpha  + \frac{1}{2}h \sin \alpha \right) h\, dx dy - \lambda \int_\Omega h\, dx dy .
\end{equation}
Formal minimization of the functional $F(h)$ with respect to $h$ leads to the following partial differential equation:
\begin{equation}
\gamma \nabla.\left(\frac{\nabla h}{\sqrt{1+(\nabla h)^2}}\right)=\rho g \left(\cos(\alpha)x-h\sin(\alpha)\right) + \lambda,
\end{equation}
with the constant contact angle condition at the wall:
\begin{equation}
    \frac{\nabla h}{\sqrt{1+(\nabla h)^2}}.\mathbf{n}=-\cos(\phi),
\end{equation}
\noindent where $\mathbf{n}$ is the unit exterior normal to the tube wall. The value of $\lambda$ can be set by integrating the resulting equilibrium equation within the whole domain, leading to the following expression:
\begin{equation}
    \lambda= - \frac{2 \gamma}{R} \cos(\phi)+ \rho g h_0 \sin \alpha,
\end{equation}
where ${h}_0=V/\Omega$ is the reference average value of the static cap height. 
By imposing $\lambda=0$, the reference height reads $h_0 = \frac{2 \ell_c^2 \cos(\phi)}{ R \sin \alpha}$, reminiscent of the well-known Jurin height. Upon non-dimensionalization with the tube radius $R$, one obtains the equilibrium equation reported in the main text Eq.\eqref{eq:YL_equation}.

\end{appendix}

\bibliographystyle{jfm}
\bibliography{Bibliography}

\end{document}